\documentclass{article}
\usepackage{fancyhdr}
\usepackage[american]{babel}
\usepackage{tikz}
\usepackage[utf8]{inputenc}
\usepackage{csquotes}
\usepackage{comment}
\usepackage{amsmath}
\usepackage{graphicx,psfrag,epsf}
\usepackage{enumerate}
\usepackage{url} % not crucial - just used below for the URL 
\usepackage{amssymb}
\usepackage{listings}
\usepackage{geometry}
\usepackage{bm, float}
\usepackage{setspace}
\usepackage{listings}
\usepackage{relsize}
\usepackage{mathtools}
\usepackage{xcolor}
\usepackage{placeins}

\newcommand{\lGamma}{\text{log}\Gamma}
\newcommand{\mean}[1]{\overline{#1}}

\begin{document}

\begin{titlepage}
    \begin{center}
        \vspace*{1cm}
            
        \large
        \textbf{Analysis of Compositional Data with Positive Correlations among Components using a Nested Dirichlet Distribution with Application to a Morris Water Maze Experiment}
            
 %       \vspace{0.5cm}
 %       \LARGE
 %       Thesis Subtitle
            
        \vspace{1.5cm}
            
        Jacob A.~Turner \\ Stephen F.~Austin State University,\\ Bianca Luedeker \\ Northern Arizona University,\\ Monnie McGee\\ Southern Methodist University
            
\end{center}

%\title{}
%\shorttitle{Nested Dirichlet Distribution}

%\leftheader{Turner, Luedeker, and McGee}
\end{titlepage}

\newpage

\begin{center}
\textbf{Abstract}
\end{center}

In a typical Morris water maze experiment, a mouse is placed in a circular water tank and allowed to swim freely until it finds a platform, triggering a route of escape from the tank. For reference purposes, the tank is divided into four quadrants: the target quadrant where the trigger to escape resides, the opposite quadrant to the target, and two adjacent quadrants. Several response variables can be measured: the amount of time that a mouse spends in different quadrants of the water tank, the number of times the mouse crosses from one quadrant to another, or how quickly a mouse triggers an escape from the tank. When considering time within each quadrant, it is hypothesized that normal mice will spend smaller amounts of time in quadrants that do not contain the escape route, while mice with an acquired or induced mental deficiency will spend equal time in all quadrants of the tank. Clearly, proportion of time in the quadrants must sum to one and are therefore statistically dependent; however, most analyses of data from this experiment treat time in quadrants as statistically independent. A recent paper introduced a hypothesis testing method that involves fitting such data to a Dirichlet distribution. While an improvement over studies that ignore the compositional structure of the data, we show that methodology is flawed. We introduce a two-sample test to detect differences in proportion of components for two independent groups where both groups are from either a Dirichlet or nested Dirichlet distribution. This new test is used to reanalyze the data from a previous study and come to a different conclusion.

\vspace{5mm}


\vfill

\newpage

\begin{center}
    \textbf{Translational Abstract}
\end{center}
The Morris water maze experiment involves a circular water tank and an escape platform hidden below the surface of the water. A mouse or rat is placed in a circular water tank and allowed to swim freely until it finds a platform. During a trial, researchers record the amount of time that a rodent swims in any one of four quadrants of the circular tank. These times are  presented as a proportion relative to the total time spent in the maze.  Researchers often want to show whether time spent in different parts of the tank differs for normal vs. memory impaired rodents. These data are usually analyzed with multiple two-sample tests to compare the proportion of time spent in each quadrant of the maze. Such an analysis treats the time measurements as statistically independent; however, these four measurements, because of their relative measure, must sum to 1. Thus, data from a water maze experiment are compositional in nature, and should be analyzed with technique developed for compositional data. We develop tests for evidence of a difference in the composition of time spent in the maze between two groups of mice. The tests we present have application to any data that are compositional in nature, and we provide code and examples for practical use.

\vfill
%\newpage
%%%%%%%%%%%%%%%%%%%%%%%%%%%%%%%%%%%%%%%%%%%%%%%%%%%%%%%%%%%%%%%%%%%%%%%%%%%%%%

\section{Introduction}
\label{sec:intro}

% Outline
% 1. Translational abstract - mention importants of WM experiments (Bianca - done)
% 2. Introduction - no heading on Intro (Monnie - done)
% 3. Overview of Compositional data analysis
%% log-ratio analysis (Monnie - done)
%% Dirichlet distribution & drawbacks (Monnie - done)
%% Nested Dirichlet distribution 
 % 4. Water maze experiments
%% Describe Maugard data and their analysis - maybe put simulation results as supplemental data and mention the issues in the body of the paper.
 % 5.  Methodology for two-sample test from Turner thesis (Jacob - done)
 % 6. Type I properties of two-sample test from Turner thesis (Jacob)
 % 7. Apply DD to Maugard data
 % 8. Explain NDD, give tree with no real explanation, give results of NDD test with Maugard data
 % 9. Future directions: tree finding algorithm, multiple groups, nested dirichlet

% Commenting out the title because the model paper does not
% have a section title for the Introduction.
%\section{Introduction %\label{sec:intro}}

\begin{figure}[h]
\centering
    \includegraphics[scale=0.1]{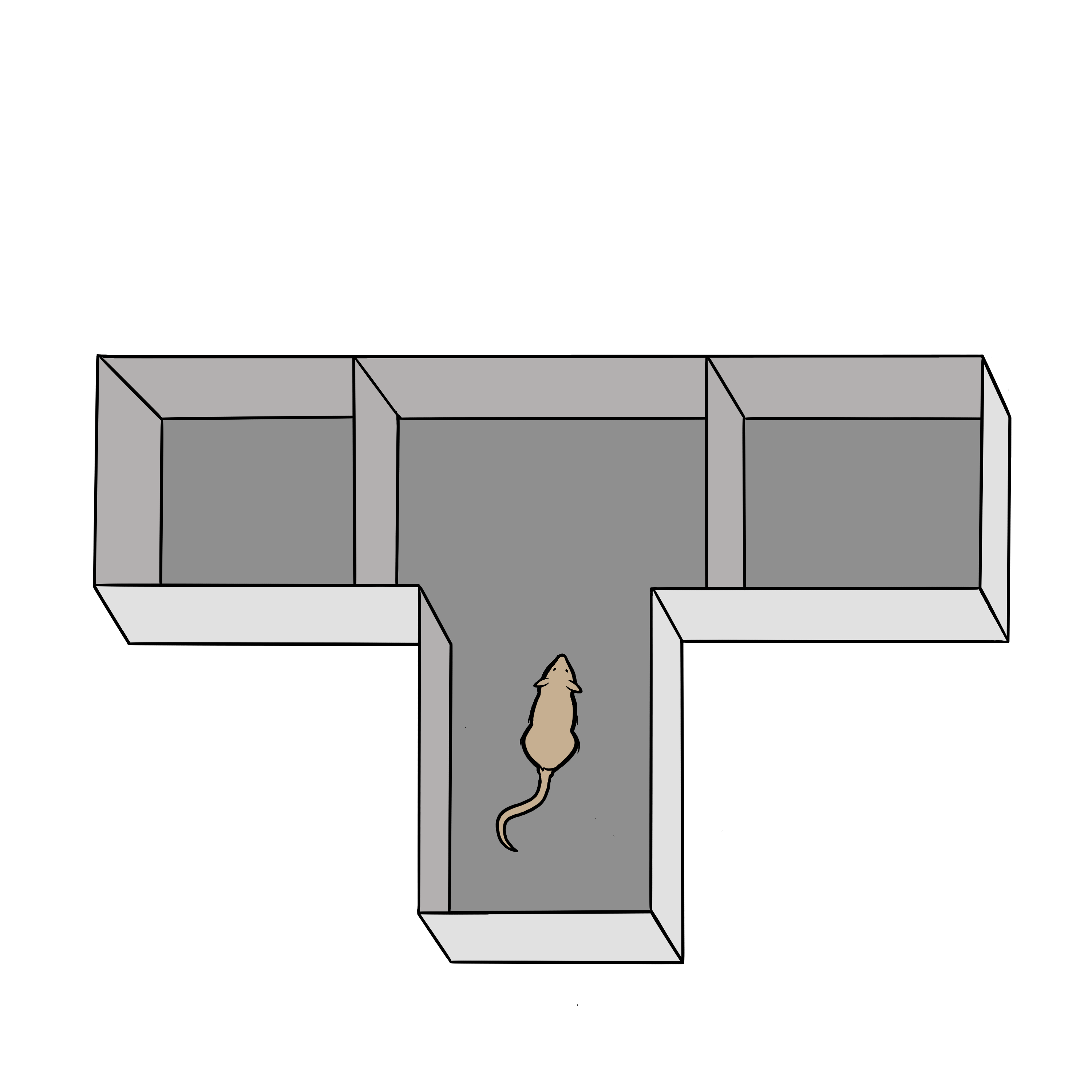}
    \includegraphics[scale=0.1]{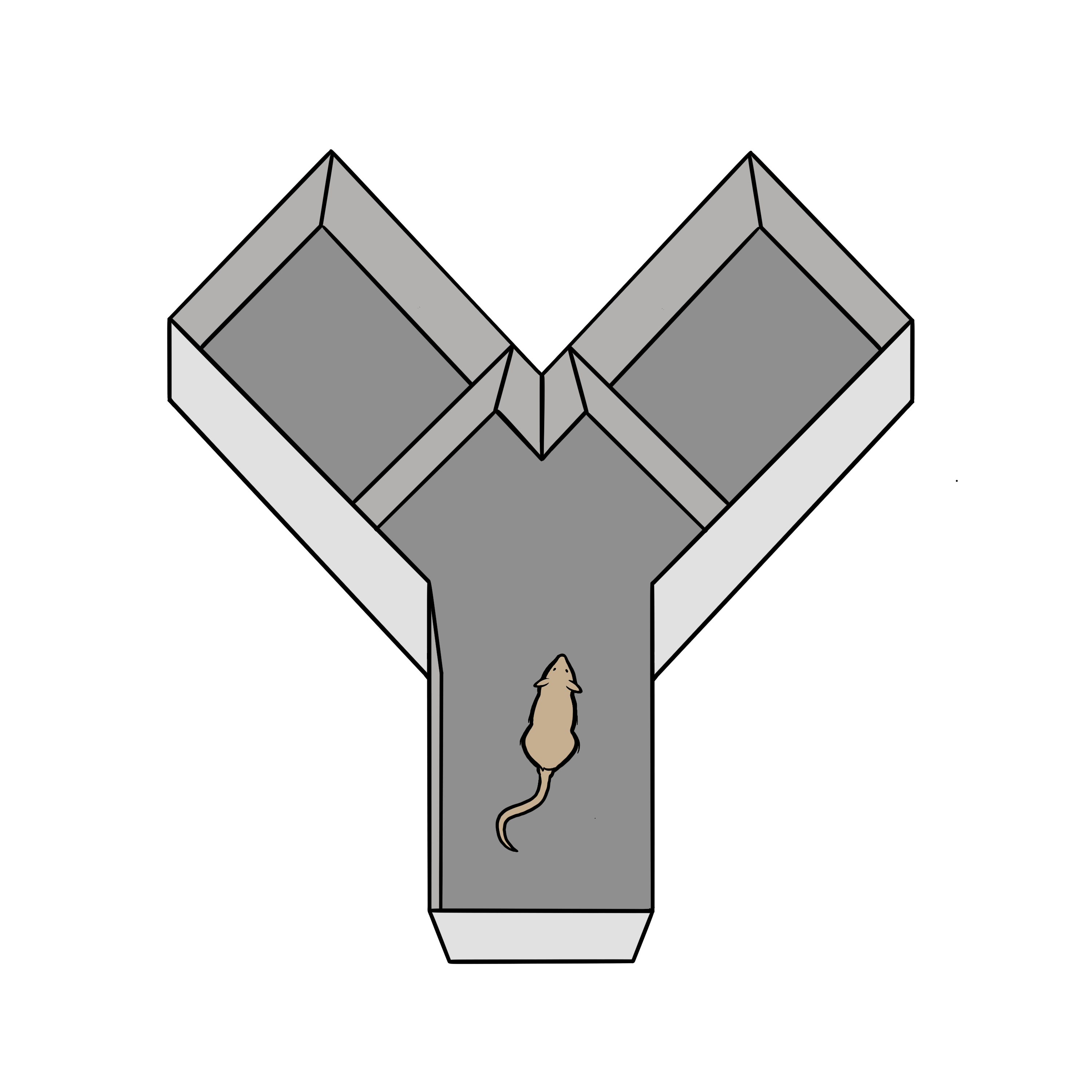}
    \includegraphics[scale=0.1]{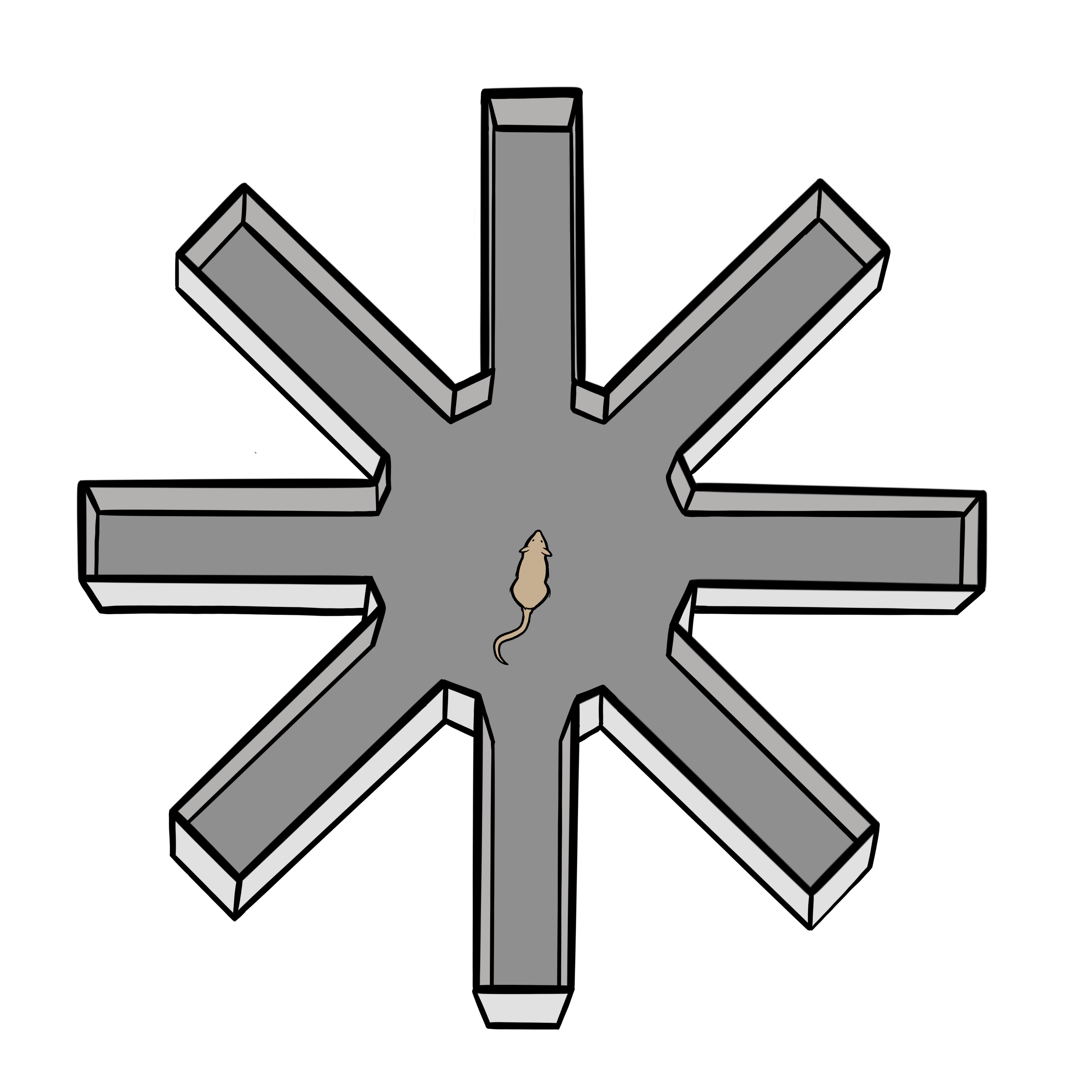}
    \caption{Three types of mazes typically used with mice and rats to assess spatial learning.  These are the t-maze, y-maze, and radial maze.  Although the rodent cannot see the goal, they are able to smell the treat at the end.  This image was based on the graphics in \cite{Leising2009AssociativeBO}.  Drawn by Calliope Luedeker.}
    \label{MiceMazes}
\end{figure}

The Morris water maze test was initially conceived to assess the spatial learning of rats \cite{morris}.  Prior to the inception of the Morris water maze, the spatial acuity of rats was evaluated using mazes where the cue to navigate the maze, such as a tasty treat at the end, was within the maze itself. Examples of these types of mazes are the t-maze, y-maze, and radial maze, shown in  Figure \ref{MiceMazes}.  The Morris water maze experiment is designed to determine whether mice and rats could recognize and remember cues outside a maze, such as a bookcase on one side of a room, to escape the maze. 

\begin{figure}[h]
\centering
    \includegraphics[scale=0.1]{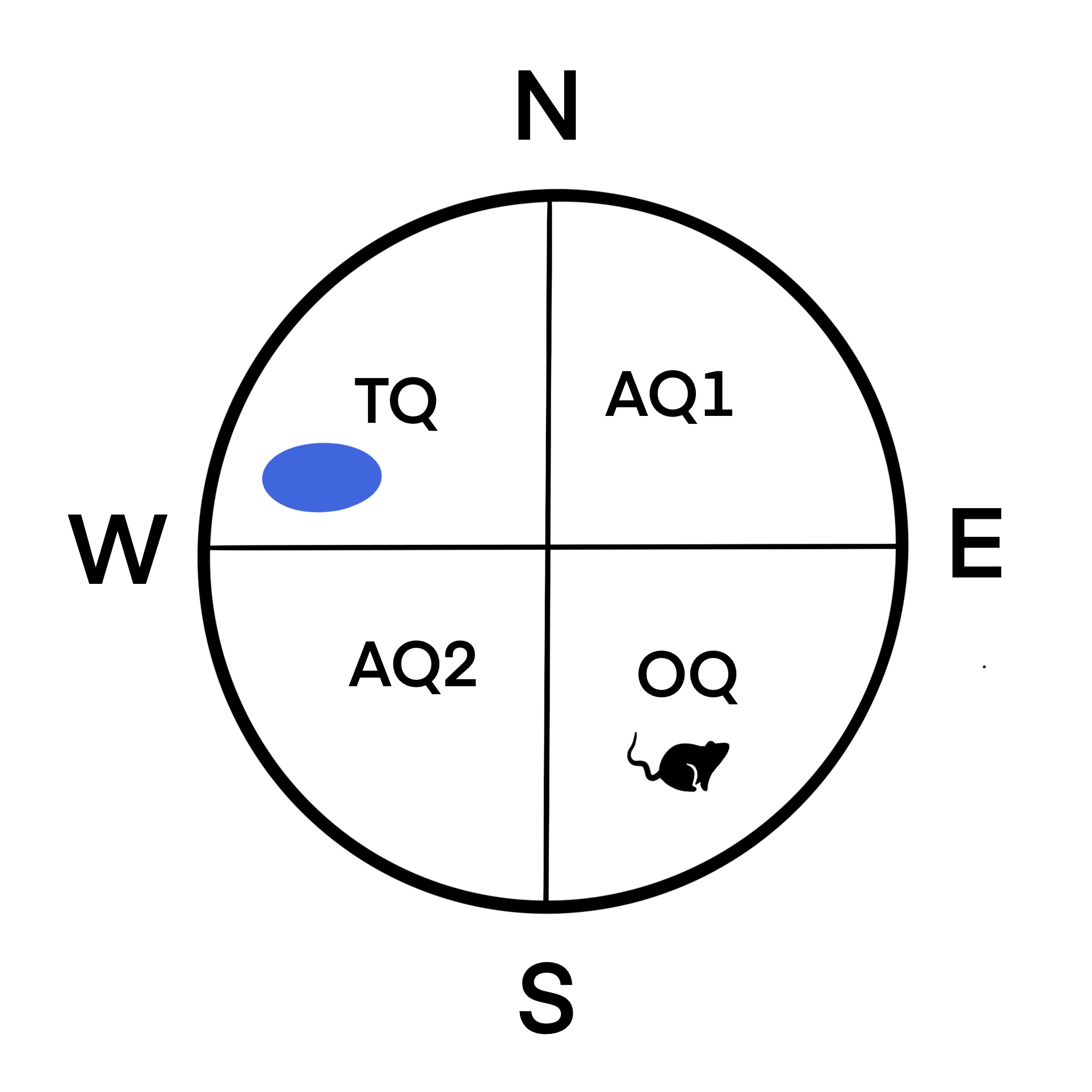}
    \caption{A diagram of a Morris water maze with cardinal directions for reference. The mouse is located in the southeast quadrant, also labeled the opposite quadrant (OQ) to where the hidden platform (solid oval) is located in the northwest quadrant. The NW quadrant is the target quadrant (TQ). The quadrants adjacent to the target are labeled AQ1 and AQ2.}
    \label{WaterMaze}
\end{figure}

To conduct a water maze experiment, mice or rats are placed in a circular tank of warm water, opacified by the addition of milk or tempera paint. Two imaginary perpendicular line segments divide the tank into quadrants as shown in Figure \ref{WaterMaze}.  The endpoints of the segments are labeled with the cardinal directions north, east, south, and west. These labels do not correspond to the actual directions; they serve only as a convenient labeling system.  In one quadrant, called the target quadrant (TQ), a hidden platform is located 1 cm below the surface of the water equidistant from the side and center of the tank and the two perpendicular lines (solid oval in NW quadrant of Figure \ref{WaterMaze}). The platform is made from clear or white plastic so that it is not visible from the viewpoint of the rodent.  Using visual cues outside the maze, the rodent can swim to the platform and use it to escape the water. Rodents are introduced into the maze over multiple trials conducted on multiple days \cite{Morris1984, Vorhees2006MorrisWM, Tian2019}. Virtual versions of the Morris water maze experiment in conjunction with fMRI imaging have been used for examining age-related spatial learning in humans \cite{zhong2017}, memory ability in an Altzheimer’s mouse model \cite{Tian2019}, and learning assessment in individuals with traumatic brain injury \cite{Andersen2020AcceleratedFT}. 
 %
%
% Commenting out these paragraphs because I don't think they are needed. MM 7-27-23
%% Experimenters have typically performed four types of experiments using the water maze: a cue test \cite{morris}, a hidden platform test \cite{Andersen2020AcceleratedFT}, a transfer test \cite{pan2011}, and a reversal test \cite{maugard}.  Each type of experiment is meant to evaluate a certain type of memory within a rodent. In addition, the water maze has been used extensively to measure the effects of drugs, brain lesions, and disease, on spatial memory and learning in the subjects.  For example, a follow-up study to the paper where the water maze is introduced explores the effect of hippocampal and cortical lesions on the spatial learning of rats \cite{Morris1984}.  
%  \cite{zhong2017} uses a virtual water maze to determine the effects of age on spatial memory and learning in humans in young adults,  high performing older adults, and low performing older adults. The results showed evidence of better performance for young adults than for either group of older adults.  A follow study, in which subjects were placed into an fMRI machine as the virtual water maze test was conducted to determine which parts of the brain were being activated, confirmed these differences  \cite{Reynolds2019}.

The response for variations on the Morris water maze test has been either occupancy--based, error--based, counting--based \cite{maei2009}, or latency--based \cite{Morris1984}.  Occupancy based measures include the percent time in the target quadrant, the quadrant of the maze where the platform is located. Error--based measures involve estimating the proximity of the mouse to the former platform location in experiments where the platform is removed or transferred to another quadrant. Counting-based methods involve counting the number of times the mouse crosses over the location of the platform (or its former location). Escape latency is the amount of time required for the rodent to find the platform and escape the maze. Typically, these data are analyzed by comparing the response variable (e.g. escape latency) across groups with an independent-samples t-test (for 2 treatment groups) or an ANOVA (for multiple groups) or with a nonparametric version of either test. 

\cite{maei2009} perform a sensitivity analysis of the various univariate measures used in 1600 mouse probe tests. They found that mean proximity to the platform location outperformed all other measures in terms of ability to detect group differences. When the measurements collected are escape latency, \cite{Tian2019} states that ANOVA is the method to use.  A newer approach is to use corrected cumulative proximity to the target as the measurement of performance in all four experiment types \cite{zhong2017, Reynolds2019}.  Realizing that the data from the cue and hidden platform tests can be censored,  \cite{Andersen2020AcceleratedFT} shows that ANOVA is not the correct approach when using escape latencies and instead uses a special type of survival model.   \cite{vouros2018} use machine learning to analyze the paths of the rodents as they swim around the maze. The paths are broken into overlapping segments of a fixed length, with each segment classified into one of nine types of behaviors by an ensemble of classifiers using majority voting.  The authors compare  a group of stressed and normal mice by conducting nine univariate Friedman tests \cite{friedman}, one for each classification of the paths.  A recent protocol on how to analyze  data from a water maze experiment prescribed that a t-test or ANOVA should be conducted on only the proportion of time in the target quadrant \cite{Vorhees2006MorrisWM}. 


The Morris water maze experiment is still used for various types of memory studies on humans and rodents even 40 years after its inception; therefore, it is important that the results from water maze studies be analyzed properly and that proper methods of analysis be widely taught and acknowledged. Using information, whether it is time spent, mean proximity, or tranversal counts, from only one of the quadrants ignores data from the other three quadrants and does not account for the necessary dependence between the four quadrants of the maze. Clearly, if a mouse spends a great deal of time in one quadrant, then it is not spending time in other quadrants. In addition, time spent in each quadrant should sum to the total amount of time in the maze \cite{maugard}. Therefore, the data are compositional. While many authors have realized the limitations of ANOVA and other previous analyses of water maze data, none have taken into account the compositional nature of the data, even when employing sophisticated techniques. 

 In the next section, we formally define compositional data and discuss previous methods, such as the use of log ratios for transforming the data \cite{aitch1982}, for analyzing data with this particular type of dependence. A recent paper appears to be the first paper to use data from all four quadrants to analyze data from a water maze experiment, thus acknowledging and incorporating the compositional nature of the data \cite{maugard}. The authors compare and contrast previous pairwise univariate methods of analysis to the Dirichlet distribution, and show the the Dirichlet distribution is truer to the structure of the data than classical methods. We introduce notation for the Dirichlet distribution; then we explain and critique the original analysis in \cite{maugard}. For comparison, we apply a statistical test using compositional analysis via a centered log ratio transform \cite{aitch1982} to the Maugard data. Following the CLR analysis, we describe a novel two-sample test for the equality of mean  vectors of components between two Dirichlet distributions. However, the Dirichlet distribution has two well-known restrictions. The first is that components with the same mean must have the same variance, and the second is that correlations between components must be negative. To address these two deficiencies, we describe the nested Dirichlet distribution \cite{dennis1991, minka1999, null2009} which generalizes the Dirichlet distribution to positive correlations and more flexible variance structures. We introduce a two-sample test for compositional data using the nested Dirichlet distribution and  apply it to the Maugard data set. We conclude with future directions for methodological development.

\section{Compositional Data}\label{sec:comp}

Compositional data are multivariate data made up of components that sum to a fixed value. Often the data are presented as proportions of a whole, where the value of each component is constrained to be between 0 and 1 and the sum of the components is 1. Such data occur naturally in a variety of applications.  The proportion of sand, silt, and clay at different depths in a lake in the Arctic \cite{coakley} logically sums to 1, indicating a dependence among proportions.  \cite{hickey2015} examined the proportions of sawlog, pallet, stake, and pulp retrieved from trees in a forest compartment. Daily physical activity data divided into sleep, physical activity, and sedentary behavior is compositional, and it is likely the relative proportions of these components, rather than the amount of time spent in any one of them, influence weight gain (or loss) in an individual \cite{gupta2018}. Finally, several high-throughput biological assays produce data that is measured in proportions of the whole. Methods designed for compositional data have recently been applied to analyze the composition of microbes in various environments \cite{koslovsky2020, tang2017} and of cell populations in flow cytometry data \cite{turner2013}. Interestingly, both physical activity data and microbial data are likely to have components that are positively correlated.

%Although recognized as common in social science and generally accepted to diminish response biases in psychological measurement, compositional measures were abandoned due to the lack of statistical methods that took into account the dependencies in the data \cite{sb2022}. Compositional data is frequently analyzed using an ANOVA or a t-test on only one component of the composition at a time \cite{aitch1982, zhong2017, hickey2015, Reynolds2019}.  Other methods of analysis include the use of if-then statements in Excel and multivariate linear regression to predict proportions of various components \cite{hickey2015}. All such methods are meant for independent variables. As such, they do not account for associations among components and they ignore the fact the data is constrained between 0 and 1. Hence, independent samples analysis on data that are compositional in nature can result in outputs that sum to more than one \cite{aitch1982, turner2013}. 

The year 2022 marked the 40th anniversary of  the first paper to address both the vast applications of compositional data and the difficulty in analysis  \cite{aitch1982, greenacre}. Before 1982, compositional data were commonly recognized in social science and generally accepted to diminish response biases in psychological measurement; however, such data were often not used due to a lack of sound methology for their anaylsis \cite{sb2022}.  Aitchison (1982) presents an additive log-ratio transformation. This transformation, given by  $\text{alr}(\bf{x}) \coloneqq [ \ln (x_1/x_k), \dots , \ln (x_{k-1}/x_k)]$, where there are $K$ components, and $x_k$, $k= 1, \ldots, K$, is the value of the response for the $k$th component. The ALR maps a vector of proportions onto $\bf{R}^{k-1}$, thus allowing standard classical or nonparametric methods to be used. However, the ALR has a drawback because the analyst must choose a reference group for the denominator of the ALR transform. For some applications, the choice of a reference group is arbitrary, and different choices lead to different outcomes. The centered-log ratio transform (CLR), given by $\text{clr}(\bf{x})\coloneqq (\ln x_i - \frac{1}{K}\sum_{j=1}^K\ln x_j)_i$ maps a K-part composition to a K-dimensional Euclidean space and does not require the choice of a reference group. The isometric log ratio (ILR) was introduced later as a transformation that preserves all metric properties, and thus allows for more complex statistical analyses \cite{egozcue}. The procedure (for any log ratio) is to transform the composition percentages and then use the transformed values in a classical analysis in place of original amounts. Once the classical analysis is complete, the data are back-transformed to obtain predicted proportions for each component. Brief comparisons of the merits and demerits of the ALR, CLR, and ILR transformations can be found in \cite{sb2022, greenacre}, and an in-depth description of a suite of log-normal transformations and other methods of analyzing compositional data in R can be found in the textbook by \cite{vanden2013}. 

Recently, there have been many studies that have moved away from the techniques introduced by Aitchison and have instead returned to methods using the Dirichlet and Dirichlet-multinomial distributions \cite{hijazi2009modeling,PackageCompositions, wang2017}.  However, the DD has an important constraint that correlations between components are nonpositive, which renders it unrealistic for data where correlations between proportions are positive.  Therefore, log ratio methods became favored in social sciences for the analysis of compositional data \cite{greenacre}. In this paper, we discuss a generalization of the DD, called the Nested Dirichlet Distribution (NDD) for analyzing Morris water maze data. We compare results from analysis of the data in \cite{maugard} using a CLR transform, a hypothesis test based on the Dirichlet distribution used in \cite{maugard}, a two-sample test for Dirichlet distributed data, and a two-sample test where the data come from a nested Dirichlet. The two-sample tests based on the Dirichlet and nested Dirichlet distributions, including their associated confidence intervals, are novel methodology for compositional data analysis.

%First, we give a summary of the experiment and the analysis from \cite{maugard}. Then, we present a critique of it and show simulation results supporting our critique. Next, we describe a two-sample test for compositional data that is based on the Dirichlet model \cite{turner2013}. We then reanalyze the data using the two-sample compositional test. Contrary to what was concluded in \cite{maugard}, use of the Dirichlet model for the data reveals that there is no evidence for a difference in time spent in the target quadrant of the maze between wild type and 3TG mice. 

\section{The Dirichlet and Nested Dirichlet Distributions }
The Dirichlet distribution is often used as the conjugate prior to the multinomial distribution in Bayesian analysis, and it has also been used to analyze compositional data \cite{ng2011}.   Suppose a composition is made up of $K$ variables, also referred to as components.   The appropriate Dirichlet distribution will have $K$ parameters $\bm{\alpha} = (\alpha_1, \dots, \alpha_k),\  k=1,\ldots, K$.  Denote this distribution as $\text{DD}(\alpha_1, \dots,  \alpha_k)$ and let $\bm{X}=[X_1, \dots X_k]^{T}$ be a random vector distributed as $\text{DD}(\alpha_1, \dots,  \alpha_k)$. Let the precision $A = \sum_{j=1}^{k} \alpha_j$. 

The density function is given by 
\begin{equation}
  f(\bm{x}|\bm{\alpha}) = \frac{\Gamma(A)}{\Gamma(\alpha_1) \dots \Gamma(\alpha_k)} \prod_{j=1}^k x_j^{\alpha_j-1} \quad  0\leq x_j \leq 1\ \text{for}\ j=1, \dots ,k.
  \label{eq:dd}
\end{equation}

The mean of a Dirichlet parameterized by \ref{eq:dd} is given by $\text{E}[X_i]=\pi_i$, and  the variance is given by:
\begin{equation}
    \text{Var}[X_i] = \frac{\pi_i(1-\pi_i)}{A+1}.
\end{equation}
Therefore, components with the same mean must also have the same variance \cite{null2009}, which limits the applicability of the DD. 

Another limitation is that the covariance between any two components is non-positive \cite{minka1999}.  Specifically:
\begin{equation}
\text{Cov}[X_i, X_j] = - \frac{\alpha_i \alpha_j}{A^2(1+A)}  = -\frac{\pi_i \pi_j}{A+1}
\label{eq:cov}
\end{equation}
Note that $\pi_i$ and $A$ are always greater than 0.

Compositional data sets seen in practice typically do not follow these constraints. This was one of Aitchison's concerns with the use of the Dirichlet distribution, and his reasoning for using various log--ratio transformations on the variables instead of applying a  Dirichlet distribution to model compositional data \cite{aitch1982}. More flexible distributions, such as the Dirichlet--multinomial \cite{mcdonald2015} and the nested Dirichlet distribution \cite{minka1999, null2008, null2009} have been proposed for the analysis of compositional data, as these distributions allow for different means and variances and for positive correlation among the components. 
%In partcular, The nested Dirichlet distribution (NDD) \cite{null2008, null2009} is a variation of the DD that allows for more complex covariance structures.

The nested Dirichlet distribution (NDD) is also called the Dirichlet-tree distribution \cite{minka1999, tang2017} and the Hyper-Dirichlet type I distribution \cite{dennis1991}. The NDD relaxes the constraints that variables with the same mean must have the same variance and that the covariance between variables is always negative \cite{null2009}.  The correlation structure between variables is determined by how the variables are nested. The nested Dirichlet contains the Dirchlet density as a special case with carefully selected parameters.

\begin{figure}
    \centering
   \begin{tikzpicture}
    \node[circle, draw, minimum size = 3em]{Root}[sibling distance= 5 cm, level distance = 2 cm]
        child {node[circle, draw, minimum size = 3em] {$N_1$} [sibling distance = 2 cm] 
            child{node[circle, draw, minimum size = 3em] {$X_1$}
            (0, -1) node{$p_1$}
            edge from parent node[left, xshift=-0.2cm] {\Large $\pi_{11}$}}
            child {node[circle, draw, minimum size = 3em] {$X_2$}
            (0, -1) node{$p_2$}
            edge from parent node[left, xshift=0.1cm] {\Large $\pi_{12}$}}
            child {node[circle, draw, minimum size = 3em] {$X_3$}
            (0, -1) node{$p_3$}
            edge from parent node[right, xshift=0.2cm] {\Large $\pi_{13}$}}
            edge from parent node[left, xshift=-0.2cm]{\Large $\pi_1$}
        }
        child {node[circle, draw, minimum size = 3em] {$N_2$} [sibling distance = 2 cm]
            child {node[circle, draw, minimum size = 3em] {$X_4$}
            (0, -1) node{$p_4$}
            edge from parent node[left, xshift=-0.2cm] {\Large $\pi_{21}$}}
            child{node[circle, draw, minimum size = 3em] {$X_5$}
            (0, -1) node{$p_5$}
            edge from parent node[right, xshift=0.2cm] {\Large $\pi_{22}$}}
            edge from parent node[right, xshift=0.2cm]{\Large $\pi_2$}
        };
    \end{tikzpicture}
    \caption{Tree diagram used to illustrate the nested Dirichlet distribution.}
    \label{fig:toy_tree}
\end{figure}
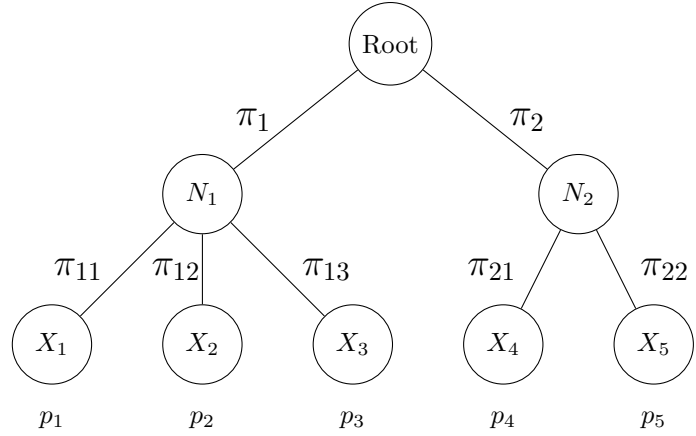

The structure of a nested Dirichlet distribution can be motivated visually with a tree diagram, as seen in Figure \ref{fig:toy_tree}.  Suppose that we have a  random compositional vector $\bm{X}=[X_1, \dots ,X_5]^{T}$ of proportions for an experiment with $K=5$ components. These variables are represented as terminal nodes in Figure \ref{fig:toy_tree}. Denote the expected value $\text{E}[X_j]= p_j$ for $j=1, \dots 5$. There are three interior nodes in the diagram, denoted $N_1, N_2$, and Root.  We say that $X_1, X_2 $ and $X_3$ are nested under $N_1$.  The last two variables, $X_4$ and $X_5$, are nested under $N_2$.  All the variables are nested under the root node although it is not their immediate parent node.  

The incorporation of interior nodes in the tree allows one to consider generating smaller compositional data sets (subtrees) based on the nesting structure defined by the tree.  For example, since $X_1, X_2$, and $X_3$ are nested underneath node $N_1$, we can create a tri-variate compositional random vector by dividing each of the three variables by their sum denoted $\bm{b_{N_1}}=[ b_{11},b_{12},b_{13}]^T=[\frac{X_1}{X_1+X_2+X_3},\frac{X_2}{X_1+X_2+X_3},\frac{X_3}{X_1+X_2+X_3}]^T$. For the variables nested under $N_2$, denote $\boldsymbol{b_{N2}}$ in a similar fashion.  $N_1$ and $N_2$ are nested underneath the root node and thus a third composition can be constructed as $\bm{b_{Root}}=[ b_{1},b_{2}]^T=[\frac{X_1+X_2+X_3}{X_1+X_2+X_3+X_4+X_5+X_6},\frac{X_4+X_5}{X_1+X_2+X_3+X_4+X_5+X_6}]^T$.  Note that in this final case, the denominator sums to one because all of the variables are nested under the root node. The expected values of the smaller compositional data sets are $E(\bm{b_{Root}})=[\pi_1,\pi_2]^T$, $E(\bm{b_{N_1}})=[\pi_{11},\pi_{12},\pi_{13}]^T$, and $E(\bm{b_{N_2}})=[\pi_{21},\pi_{22}]^T$ and are displayed along the branches in Figure \ref{fig:toy_tree}. 

\begin{figure}
    \centering
   \begin{tikzpicture}
    \node[circle, draw, minimum size = 3em]{Root}[sibling distance= 5 cm, level distance = 2 cm]
        child {node[circle, draw, minimum size = 3em] {$N_1$} [sibling distance = 2 cm] 
            child{node[circle, draw, minimum size = 3em] {$X_1$}
            edge from parent node[left, xshift=-0.2cm] {\Large $\alpha_1$}}
            child {node[circle, draw, minimum size = 3em] {$X_2$}
            edge from parent node[left, xshift=-0.0cm] {\Large $\alpha_2$}}
            child {node[circle, draw, minimum size = 3em] {$X_3$}
            edge from parent node[right, xshift=0.2cm] {\Large $\alpha_3$}}
            edge from parent node[left, xshift=-0.2cm]{\Large $\alpha_6$}
        }
        child {node[circle, draw, minimum size = 3em] {$N_2$} [sibling distance = 2 cm]
            child {node[circle, draw, minimum size = 3em] {$X_4$}
            edge from parent node[left, xshift=-0.2cm] {\Large $\alpha_4$}}
            child{node[circle, draw, minimum size = 3em] {$X_5$}
            edge from parent node[right, xshift=0.2cm] {\Large $\alpha_5$}}
            edge from parent node[right, xshift=0.2cm]{\Large $\alpha_7$}
        };
    \end{tikzpicture}
    \caption{Tree diagram relabeled with $\alpha$ parameters}
    \label{fig:alpha_tree}
\end{figure}
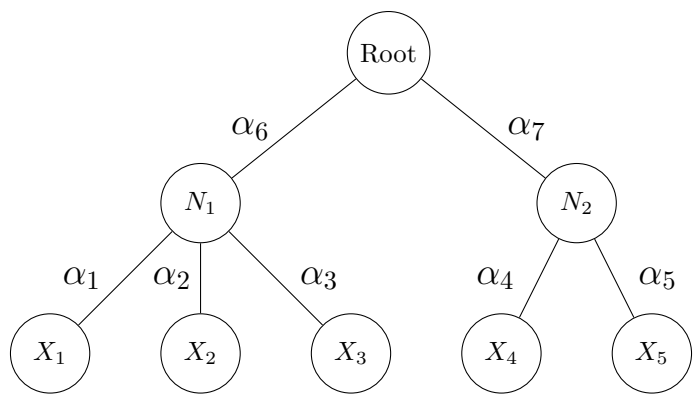

The nested Dirichlet distribution can be derived by assuming that each of the individual subtrees are independently distributed Dirichlet random vectors.  The mapping of the random variables within each subtree to the compositional vector $\boldsymbol{X}$ is a one-to-one transformation and thus, standard transformation theory applies.  In our example, assume that $\boldsymbol{b_{root}}\sim DD(\alpha_6,\alpha_7)$, $\boldsymbol{b_{N_1}}\sim DD(\alpha_1,\alpha_2,\alpha_3)$, and $\boldsymbol{b_{N_2}}\sim DD(\alpha_4,\alpha_5)$.  Figure \ref{fig:alpha_tree} illustrates the correspondence between the parameters and their location on the tree diagram.  

The joint density of $\boldsymbol{X}$ is a nested Dirichlet density, obtained via transformation, and is proportional to, up to a normalizing constant,
\begin{equation}
  f(\bm{x}|\bm{\alpha}) \propto  \prod_{j=1}^k x_j^{\alpha_j-1} \left( \sum_{j=1}^3 x_j \right)^{\alpha_6-\sum_{j=1}^3 \alpha_j} \left( \sum_{j=4}^5 x_j \right)^{\alpha_7-\sum_{j=4}^5 \alpha_j} \quad  0\leq x_j \leq 1\ \text{for}\ j=1, \dots ,k.
  \label{eq:ndd}
\end{equation}
where $\boldsymbol{\alpha}$ is the vector of parameters $(\alpha_1,\alpha_2,...,\alpha_7)^T$.  It is easily verified that if $\alpha_6=\sum_{j=1}^3 \alpha_j$ and $\alpha_7=\sum_{j=4}^5 \alpha_j$, then the joint density reduces to the standard Dirichlet density. From the tree diagram perspective, if the parameter associated with an internal node is equal to the sum of immediate children's parameters, then the node is removed from the tree.

The expected value of each component within $\boldsymbol{X}$ is easily obtained due to the properties of the standard Dirichlet and the independence of each subtree.  For example, $p_1=E(X_1)=E(b_1 b_{11})=E(b_1)E(b_{11})=\pi_1\pi_{11}$.  The relationships listed below follow similarly:
\begin{equation}
    \begin{matrix*}[l]
    p_1 = \pi_1\pi_{11}\\
    p_2 = \pi_1\pi_{12}\\
    p_3 = \pi_1\pi_{13}\\
    p_4 = \pi_2\pi_{21}\\
    p_5 = \pi_2\pi_{22}.
    \end{matrix*}
\end{equation}
We also have that
\begin{equation}
    \begin{matrix*}[l]
        \pi_1 = p_1+p_2+p_3  & \pi_2 = p_4 + p_5\\
        \pi_{11}+\pi_{12}+\pi_{13} = 1 & \pi_{21}+\pi_{22}= 1 \\
     \pi_{1j} = \frac{p_j}{\pi_1} \ \text{for} \ j= 1, 2, 3 &  \pi_{21} = \frac{p_4}{\pi_2}, \ \pi_{22} = \frac{p_5}{\pi_2}.
    \end{matrix*}
\end{equation}
Depending on the values of $\boldsymbol{\alpha}$ and the structure of the tree, two variables with the exact same mean can have potentially different variances. This added flexibility occurs when comparing two variables that are not nested under the same node. Additionally,  positive correlations are also possible and occur when the sum of the parameters for a given subtree is larger than the parameter of the subtree's parent node. In our example derivation displayed in Equation \eqref{fig:alpha_tree}, variables $X_1,X_2,$ and $X_3$ would be positively correlated if $\alpha_6 < \Sigma_{j=1}^3 \alpha_j$. For a technical derivation for the covariance and thus the correlations for all pairwise variables and any given tree, see \cite{dennis1991}. 

It is straight forward to estimate the parameters of both the Dirichlet and nested Dirichlet distributions by way of maximum likelihood estimation.  For the Dirichlet case, parameter estimates can be obtained using the method in \cite{minka2000} or via Dirichlet regression \cite{DirichletReg}.  In the case of the nested Dirichlet distribution, parameter estimates are obtained by decomposing the original compositional data into smaller subtrees defined by the distribution's tree structure and using standard Dirichlet estimation techniques for each subtree \cite{null2009}.  Additional details and discussion are presented in later sections.     

\begin{figure}
\centering
\begin{tikzpicture}
    \node[circle, draw, minimum size = 4em]{Root}[sibling distance= 5 cm, level distance = 3 cm]
        child {node[circle, draw, minimum size = 4em] {$N_1$} 
        [sibling distance = 2.5 cm] 
            child{node[circle, draw, minimum size = 4em] {AQ1}
            edge from parent node[left, xshift=-0.2cm] {\Large 11.6}}
            child {node[circle, draw, minimum size = 4em] {OQ}
            edge from parent node[right, xshift=0.2cm] {\Large 10.3}}
            edge from parent node[left, xshift=-0.2cm] {\Large 8.1}
        }
        child {node[circle, draw, minimum size = 4em] {$N_2$} [sibling distance = 2.5 cm]
            child {node[circle, draw, minimum size = 4em] {AQ2}
            edge from parent node[left, xshift=-0.2cm] {\Large 5.6}}
            child{node[circle, draw, minimum size = 4em] {TQ}
            edge from parent node[right, xshift=0.2cm] {\Large 9.2}}
            edge from parent node[right, xshift=0.2cm] {\Large 11.2}
        };
    \end{tikzpicture}
 \caption{The best fitting tree with the branches labeled with the corresponding MLEs of the $\alpha$ parameters.}
    \label{best_tree_mle}
\end{figure}
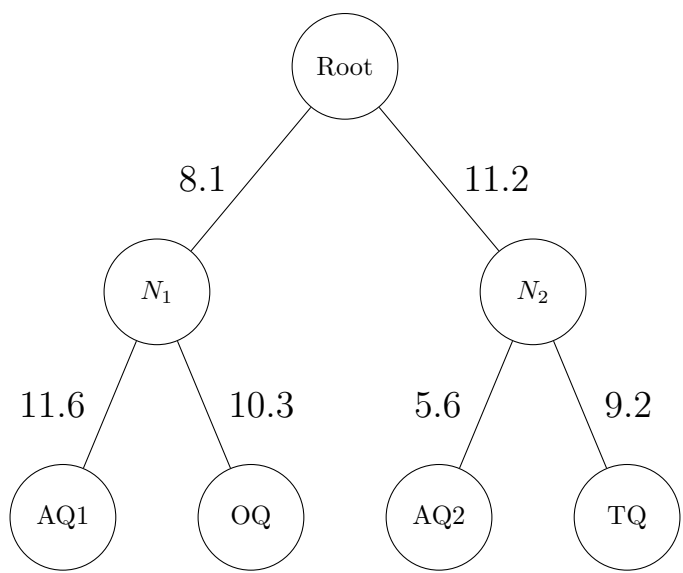

\section{Analysis of the Maugard water maze data}

\cite{maugard} examined measurements taken from performance of seven wild-type (WT) mice and seven 3xTG-AD (AD) mice in a Morris water maze. The WT  mice can be assumed to be healthy mice and the AD mice have three mutations related to Alzheimer's disease that affect the hippocampus and cerebral cortex. The goal for the Maugard experiment is to determine if the proportion of time spent in each of the four quadrants differs between the healthy mice and treatment mice. Recall that the four quadrants of a water maze are the target quadrant containing the escape platform (TQ), the quadrant opposite the target (OQ), and two quadrants adjacent to the target (AQ1 and AQ2). In this section, we explore their data using ternary diagrams, describe their hypothesis testing scenario, reproduce results of the analysis in \cite{maugard}, and discuss some logical inconsistencies with their method.

\subsection{Ternary Diagrams for Exploring Compositional Data}
\begin{figure}
    \centering
    \includegraphics[scale = 0.7]{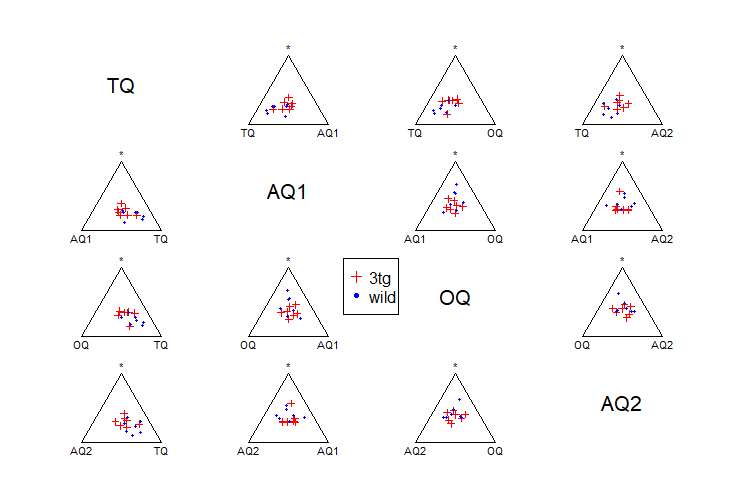}
    \caption{All possible ternary diagrams comparing the control and treatment mice.  The star at the peak of each triangle is the sum of the two components not labeled.  The blue dots represent the control group and the red crosses represent the treatment group.  There does not appear to be two distinct groups in the diagrams.}
    \label{mice_ternary}
\end{figure}

Ternary diagrams plot values of observations within a triangle to show relationships among variables whose values sum to a constant. They can aid in visualizing differences between the two groups \cite{vanden2013}, even when each group has more than three components. Each triangle represents a sub-composition with three components: the two components represented by the labeled corners and a third that is the sum of the proportions of the remaining $k-2$ components. The closer a point is to one of the corners, the higher the proportion of that component for that observation.  A point directly in the center would represent a value of $(1/3,\ 1/3,\ 1/3)^{T}$.  Figure \ref{mice_ternary} shows a matrix of all possible ternary diagrams for the Maugard data.

For example, consider the top left triangle in Figure \ref{mice_ternary}.  This triangle has the left point labeled AQ1, right point labeled TQ, and top point labeled with a star.  The star represents the sum of OQ and AQ2. A dot as a plotting symbol represents a value from the WT group, and a cross represents a value from the AD group. Focusing on the dot at the center bottom of the triangle, we see that it is far away from the star. Thus the sum of the proportion of combined time spent in OQ and AQ2 is small for this rodent, whereas it spends equal amounts of time in quadrants AQ1 and TQ.  This dot corresponds to the observation $(\text{TQ}=0.42,\ \text{AQ1}=0.35,\ \text{OQ}=0.15,\ \text{AQ2}=0.08)$.  The combined proportion of time spent in OQ and AQ2 is 0.23.  This is the smallest out of all the observations.  The red cross that is closest to the top of the triangle corresponds to the observation $(0.21,\ 0.22,\ 0.24,\ 0.33)$. The combined proportion of time spent in OQ and AQ2 is 0.57.

If there was a difference between the two groups, we would expect to see two distinct clusters in at least some of the ternary diagrams.  In this case,  there is not a clear separation between the control and treatment group in any of the diagrams.  However, one can argue that the control group (WT) has some observations closer to the target quadrant corner while the treatment group (AD) observations tend to stay clustered in the center of the diagrams. Although there is not a clear visual separation between observations from different treatments, it is still possible that there could be a real difference between treatments because the diagrams do not display all facets of the relationships among the groups. We are limited to investigating sub-compositions with three components.  This limitation could mask important relationships.

\subsection{Hypothesis Test System in Maugard (2019) \label{sec:maugard}}

%\cite{maugard} used the water maze to examine measurements taken from performance of seven wild-type (WT) mice and seven 3xTG-AD (AD) mice. The WT  mice can be assumed to be healthy mice and the AD mice have three mutations related to Alzheimer's disease that affect the hippocampus and cerebral cortex.   
In order to take advantage of the compositional structure in the water maze experiment, \cite{maugard} model their data with two Dirichlet distributions, one for each treatment group.  Specifically, let $\bm{\pi_{1}} = (\text{TQ}_{1}, \text{ AQ1}_{1}, \text{ OQ}_{1}, \text{ AQ2}_{1})^{T}$ be population mean proportion of time spent in each quadrant for WT mice and $\bm{\pi_{2}}= (\text{TQ}_{2}, \text{ AQ1}_{2}, \text{ OQ}_{2}, \text{ AQ2}_{2})^{T}$ be the same vector of parameters for AD mice.  Using this notation, they set up the following system of hypothesis tests:
\begin{equation}
\label{eqn:maugardtest1}
\begin{split}
\text{H}_0: & \hspace{0.1in} \text{TQ}_1 = \text{AQ1}_1 = \text{AQ2}_1 = \text{OQ}_1 = 0.25\\
\text{H}_1: & \hspace{0.1in} \text{At least one of the above is not } 0.25
\end{split}  
\end{equation}
\begin{equation}
\label{eqn:maugardtest2}
\begin{split}
\text{H}_0: & \hspace{0.1in} \text{TQ}_{2} = \text{AQ1}_{2} = \text{AQ2}_{2} = \text{OQ}_{2} = 0.25\\
\text{H}_1: & \hspace{0.1in} \text{At least one of the above is not } 0.25.
\end{split}  
\end{equation}

\noindent where $TQ_i$, $AQ1_i$, $AQ2_i$, and $OQ_i$, $i = 1, 2$, are the proportions of time spent in the target quadrant, adjacent quadrant 1, adjacent quadrant 2, and opposite quadrant, respectively, where $i$ indexes the treatment group (WT or AD). 

The authors posit that if the mice cannot remember where the platform was, they will show no preference for any of the quadrants. In other words, the mice will spend one quarter of the time in each quadrant. We would expect the AD mice to spend less time in the target quadrant than the WT mice because the AD mice would have difficulty remembering the placement of the platform. Likewise, the WT mice would spend more time in the target quadrant because they have a memory of the location of the target platform.

The authors obtain separate maximum likelihood estimates of  $\text{DD}(\alpha_1, \alpha_2, \alpha_3, \alpha_4)$ for the control data and the treatment data using the formula found in \cite{minka2000}.  They conduct separate hypothesis tests, one for the WT data and the other for the AD data. Their reasoning for the two tests is that they expect to reject the null for the WT mice and fail to reject the null for the AD mice. If the authors reject the null hypotheses in Equation \ref{eqn:maugardtest1} and fail to reject the  the null hypothesis in Equation \ref{eqn:maugardtest2}, they conclude that there is evidence of a difference in the amount of time spent in each quadrant between the control and treatment groups. 

Ultimately, the authors reject $H_0$ for the WT mice (Equation \ref{eqn:maugardtest1}, $p = 0.0021$) and do not reject $H_0$ for the AD mice (Equation \ref{eqn:maugardtest2}, $p = 0.26$).  They use these results to conclude that there is evidence that the the mice in the AD group spend equal amount of time in each quadrant, which indicates evidence that the AD mice have worse memories than the WT mice. 

%The authors perform a test for differences between two treatment groups by first estimating parameters for a Dirichlet distribution separately for each group, and then comparing these parameter estimates to a uniform distribution. Details on the test are given in Section \ref{sec:previous}. 
\subsection{Limitations of the Procedure}
Fitting a Dirichlet distribution is a vast improvement over using Student's t-test or ANOVA for comparing the mean amount of time spent in one quadrant across all groups; however, there are still some conceptual issues in the analysis, particularly in the way that the the authors test for differences between groups. For example, it is unclear what the authors would conclude if they fail to reject the null hypothesis for both tests. Similarly, it is unclear what the authors would conclude if they reject the null for both Equations \ref{eqn:maugardtest1} and \ref{eqn:maugardtest2}. In addition, it is not explicitly stated, but we assume that \cite{maugard} believe that Pr(Type I) for the overarching test in the manner they conducted it (Equations \ref{eqn:maugardtest1} and \ref{eqn:maugardtest2}) is 5\% as long as the hypothesis tests for the individual groups were conducted at the 5\% level.  

We conducted a simulation study to determine whether the overall Pr(Type I) error is close to the nominal value of 5\%. Note that we use Pr(Type I) to denote the probability of Type I error rather than $\alpha$ due to the use of $\alpha$ as a parameter for the DD. To estimate Pr(Type I) error, we drew 14 observations from a $\text{DD}(\alpha_1, \alpha_2, \alpha_3, \alpha_4)$ distribution using five different sets of $\alpha$ parameters given in Table \ref{water_maze_sim_results}.  Seven observations were randomly assigned to the control group and the remaining seven were assigned  to the treatment group for each set of parameters. The test outlined in Equations \ref{eqn:maugardtest1} and \ref{eqn:maugardtest2} was then applied to simulated samples. We recorded the number times that the null hypotheses for both groups were rejected (Reject Both), the number of times both hypotheses were not rejected (Fail to Reject Both), and the number of times one null hypothesis was rejected and the other was not without distinguishing which null was rejected (Reject One).  This simulation was repeated 10,000 times for each set of $\alpha$ parameters.  Because \cite{maugard} never specified what to do in the case where both null hypotheses are rejected or both are not rejected, we decided that we would only declare evidence of unequal proportions in the case where one is rejected and one is not.  The results are shown in Table \ref{water_maze_sim_results}.

\begin{table}
\caption{The simulation results using the method presented in the water maze paper. The $\alpha$ parameters for the Dirichlet distributions are given in the first column. The first row of $\alpha$ parameters are the MLEs from the Maugard study. The numbers in the other columns are the number of times out of 10,000 simulations that the event at the top of the column occurred. \label{water_maze_sim_results}}
\begin{center}
\begin{tabular}{rrrrr}
  \hline
 $\bm{\alpha}$ & Reject Both & Fail to Reject Both & Reject One & Pr(Type I) \\ 
  \hline
 (9.8, 6.1, 5.4, 5.9) & 5501 & 663 & 3836 & 0.3836 \\ 
   (6.8, 6.8, 6.8, 6.8) & 26 & 9033 & 941 & 0.0941 \\ 
   (9, 6, 6, 6) & 2307 & 2675 & 5018 & 0.5018 \\ 
  (8, 7, 7, 5) & 1812 & 3269 & 4919 & 0.4919 \\ 
  (12, 5, 5, 5) & 9918 & 0 & 82 & 0.0082 \\ 
   \hline
\end{tabular}
\end{center}
\end{table}

From the table of results, it is clear that Pr(Type I) is different from 5\%.  Instead, Pr(Type I) ranges from less than 1\% to 50\%. Furthermore, the Pr(Type I) is dependent upon the parameter values of the Dirichlet distribution.  The first set of parameters, (9.8, 6.1, 5.4, 5.9) are the maximum likelihood estimates from the set of all 14 observations (7 for each group) in the Maugard data rounded to one decimal place.  These were calculated using the \texttt{dirichlet.mle} function in the \texttt{sirt} pacakge \cite{sirt}. The  Type I error in this case represents declaring that the two groups have different mean vectors when in fact they do not. For the case given in \cite{maugard}, a Type I error occurs almost 40\% of the time.  

The Pr(Type I) is reduced in two cases. The first case is when the parameters of the Dirichlet distribution are equal (second row of Table \ref{water_maze_sim_results}).  In the simulation study, a uniform value of 6.8 was chosen to keep the precision $A = \sum_{i=1}^4 \alpha_i$ comparable between sets of parameters.  Since the distribution is uniform, we are more likely to fail to reject both null hypotheses. The other case where Pr(Type I) is low is when the MLE of one component is much different from that of the otehrs, represented by the simulation scenario where the parameter vector  is (12, 5, 5, 5).  Because the parameter vector is quite different from a uniform distribution, the Type I error is low and the power is high. Even so, the procedure in Equations \ref{eqn:maugardtest1} and \ref{eqn:maugardtest2} is not a valid procedure due to the large range of Type I errors and the logical inconsistencies. 

\section{A Two Sample Test for Equality of Dirichlet Components \label{sec:twosample}}

One might ask why \cite{maugard} used two sets of hypotheses to come to a conclusion about their experiment when they should have used a two-sample test, which would test the hypothesis in Equation \ref{mice_hypo}.
\begin{equation}
\label{mice_hypo}
\begin{split}
\text{H}_0: & \hspace{0.1in} \bm{\pi_{1}} = \bm{\pi_{2}}\\
\text{H}_1: & \hspace{0.1in} \bm{\pi_{1}} \neq \bm{\pi_{2}}
\end{split}    
\end{equation}
\noindent where $\bm{\pi_{i}},\ i = 1, 2$, represents the vector of proportions for each group. They were likely unaware of previously unpublished work deriving a two-sample test for equality of mean vectors between two Dirichlet distributions \cite{turner2013}. We introduce this test for \ref{mice_hypo} using the general framework of Dirichlet regression \cite{hijazi2009modeling}.

Let the data be drawn from a $\text{DD}(\alpha_1, \alpha_2, \alpha_3, \alpha_4)$, as given by Equation \ref{eq:dd}. Let $A_1$ be the precision parameter for treatment 1 and $A_2$ be the precision parameter for treatment 2. Let $n_1$ and $n_2$ be the respective sample sizes. Under the null hypothesis, there is a common mean vector $\bm{\pi}$.  The precision parameters for each population are allowed to vary.  If not, we would be testing not just that the means are the same, but also that the two distributions are identical, which is more restrictive than necessary.    

The log-likelihood function under the null hypothesis is:
\begin{equation}
\label{NullLikelihood}
\begin{split}
    L_0(A_1, A_2, \bm{\pi} | \text{data}) &= n_1 \lGamma(A_1) - n_1 \sum_{j=1}^{k} \lGamma(A_1\pi_j) + n_1 \sum_{j=1}^{k}(A_1\pi_j -1) \mean{\text{log}x_j}\\ 
    &+ n_2\lGamma(A_2) - n_2 \sum_{j=1}^{k} \lGamma(A_2\pi_j) + n_2 \sum_{j=1}^{k}(A_2\pi_j -1) \mean{\text{log}y_j},
\end{split}
\end{equation}

\noindent where $x_{ij}, \ i= 1, \dots, n_1, \ j= 1 \dots k$ is the observed compositional data, with $j$ components and sample size $n_1$ for treatment 1. The quantity $\mean{\text{log}x_j} = \frac{1}{n_1} \sum_{i=1}^{n_1} \text{log} x_{ij}$ for $j = 1 \dots k$ is the average log proportions for each component, $j$, in the composition. Let $y_{ij}$ and $\mean{\text{log}y_j}$ be defined similarly for treatment 2. The log-likelihood function under the alternative hypothesis is:
\begin{equation}
\label{AlternativeLikelihood}
\begin{split}
    L_1(A_1, A_2, \bm{\pi_1}, \bm{\pi_2} | \text{data}) &= n_1\lGamma(A_1) - n_1 \sum_{j=1}^{k} \lGamma(A_1\pi_{1j}) + n_1 \sum_{j=1}^{k}(A_1\pi_{1j} -1) \mean{\text{log}x_j}\\ 
    &+ n_2\lGamma(A_2) - n_2 \sum_{j=1}^{k} \lGamma(A_2\pi_{2j}) + n_2 \sum_{j=1}^{k}(A_2\pi_{2j} -1) \mean{\text{log}y_j}.
\end{split}
\end{equation}

\noindent Using these likelihoods, the likelihood ratio test statistic is
\begin{equation}
    \Lambda = -2[\max L_0(A_1, A_2, \bm{\pi} | \text{data}) 
    -\max L_1(A_1, A_2, \bm{\pi_1}, \bm{\pi_2} | \text{data})].
    \label{eq:lrttest}
\end{equation} 
  The number of free parameters under the null hypothesis  is $k+1$ and the number of free parameters under the alternative is $2k$. Taking their difference yields $k-1$. Thus, the test statistic $\Lambda$ in Equation \ref{eq:lrttest} follows an approximate $\chi^2$ distribution with $k-1$ degrees of freedom \cite{casella2002statistical}.

There is no closed form solution for the MLEs using $L_0$ or $L_1$.  Instead, a numerical method must be used to maximize the functions.  We will briefly discuss the general linear model framework of Dirichlet regression implemented in the \texttt{DirichletReg} package of the open-source statistical software suite R \cite{hijazi2009modeling, DirichletReg,rlanguage}. The log-likelihood for a single sample of $n$ observations from a Dirichlet distribution, denoted $L(\pi,A|data)$, is 
\begin{equation}
\label{LogLikereg}
\begin{split}
    L(\bm{\pi},A| \text{data}) &= n\lGamma(A) - n\sum_{j=1}^{k} \lGamma(A\pi_{j}) + n \sum_{j=1}^{k}(A\pi_{j} -1) \mean{\text{log}x_j}.\\  
\end{split}
\end{equation}
where $x_{ij}$, $i=1,...,n$, $j=1,...,k$ is the sample data and the quantity $\mean{\text{log}x_j}=\frac{1}{n}\Sigma_{i=1}^n log x_{ij}$ for $j=1,...,k$.

Dirichlet regression allows for covariates, $z_1,z_2,...,z_p$, to be included in the Dirichlet model to explain trends in the mean and account for heteroskedacity through the precision parameter \cite{DirichletReg}.  The covariates are incorporated into the model by using separate link functions for the means and precision:
\begin{equation}
\label{Linkfunctions}
\begin{split}
    g_{\bm{\pi}}(\pi_j) &= \beta_{0j}+\beta_{1j}z_1+...+\beta_{pj}z_p\\
    g_A(A) = log(A) &=\beta_{0A}+\beta_{1A}z_1+...+\beta_{pA}z_p
\end{split}
\end{equation}

The link function for the mean vector, $g_{\bm{\pi}}$, is the the multinomial logistic function to ensure that each prediction made on the mean scale sums to one.  To do this, one of the mean components serves as a reference which we will refer as the last component, $\pi_k$.  For this component, all of the regression coefficients are set to 0.  Estimates for the regression coefficients are obtained by maximizing Equation \eqref{LogLikereg} substituting $\bm{\pi}$ and $A$ with 
\begin{equation}
\label{meanlink}
\begin{split}
    \pi_j &= \frac{e^{\beta_{0j}+\beta_{1j}z_1+...+\beta_{pj}z_p}}{\Sigma_{l=1}^k e^{\beta_{0l}+\beta_{1l}z_1+...+\beta_{pl}z_p}}\\
    A &= e^{\beta_{0A}+\beta_{1A}z_1+...+\beta_{pA}z_p}
\end{split}
\end{equation}
It should be noted that it is not required that the same set of predictors be used to model the mean and precision components.  A subset of the predictors can be used to model the means, while a separate unique set of predictors can be used to model changes in the precision parameter.

The maximization of Equation \eqref{LogLikereg} with respect to the regression coefficients defined Equation \eqref{meanlink} is accomplished by implementing a three-step approach \cite{DirichletReg}. Briefly, good initial starting values are obtained by modeling each individual component using Beta regression \cite{cribari2010beta}.  With initial values obtained, two rounds of numerical optimization are applied to ensure convergence along with stable standard error estimates \cite{DirichletReg}. 

For the two sample comparison setting, a single, dummy variable covariate, $z_1$, indicating treatment status is used. To maximize $L_0$ under the null case that the mean vectors are equal to each other, one includes an intercept-only model for the means, $g_{\pi}(\pi_j)=\beta_{0j}$.  To allow for potentially different precision parameters, we let $g_A(A)=\beta_{0A}+\beta_{1A}z_1$.  To maximize the alternative case $L_1$, we specify $g_{\pi}(\pi_j)=\beta_{0j}+\beta_{1j}$ to allow for different means between the two groups while keeping the same model specification for $g_A(A)$.  

%\textcolor{cyan}{{\it Replace the following with text from dissertation.} This can be done in a variety of ways. One approach is suggested by \cite{turner2013} which estimates the precision parameters holding the mean parameters fixed and then estimates the mean parameters holding the precision fixed. The process is continued until a convergence criterion is met.} %An alternative approach is to estimate the parameters using Dirichlet regression as discussed in \cite{hijazi2009modeling} and can be implemented using the \texttt{DirichletReg} package in R version 4.1.2 \cite{DirichletReg,rlanguage}. Under the Dirichlet regression setting, a general linear model framework is used with a logit-link function for the mean parameters and log-link function for the the precision parameter.  To maximize $L_1$, one simply includes an intercept and a single covariate that differentiates the two groups. To maximize $L_0$, an intercept only model is used for the mean parameters while including an intercept and group covariate for the precision parameter to allow for unequal precision values for each group.   

Using \texttt{DirichletReg} to estimate the parameters of \eqref{LogLikereg}, we applied the two sample test for the mean of the Dirichlet model to the Maugard data set. The unresticted MLEs for the WT group are $\bm{\hat{\pi}_1}=(.423,.194,.181,.202)$ and $\hat{A}_1=27.025$.  The unrestricted MLEs for the AD group are  $\bm{\hat{\pi}_2}=(.301,.255,.216,.228)$ and $\hat{A}_2=41.678$.  In terms of point estimates, we can see a considerable shift in the mean of the first compositional variable, representing the percentage of time spent in the target quadrant (TQ). The difference between WT and AD for this component is positive (.122), while the difference for the others is much smaller and negative. However, the computed likelihood ratio test statistic is 7.149 with a corresponding p-value of .067. With this p-value, many researchers would state that there is not enough evidence of a difference in time spent among the quadrants between the two treatment groups. Therefore we come to the opposite conclusion given in the original paper. The parameter estimates give some insight as to why the p-value, while greater than 0.05, is still quite low.  The WT mice seem to be spending more time in the target quadrant than the mice in the AD group, as hypothesized; however, the change is not large enough to be detected with such a small sample size, indicating lack of statistical power. We investigate Type I and Type II errors of the test given in \eqref{eq:lrttest} in the next section. 

\subsection{Type I Error and Power for LRT}
An important distinction between the test given by \eqref{eq:lrttest} versus that of \cite{maugard} is the test's performance in terms of type I error control.  Table \ref{water_maze_sim_results2} provides type I error estimates using the likelihood ratio test for equal means for the five scenarios used in our initial investigations of the \cite{maugard} testing procedure.  In this simulation study, we also examined the performance of the two-sample Dirichlet component test in terms of sample size.  All scenarios had an equal sample size among the two groups.  As expected, the type I error rates are somewhat inflated for small sample sizes but approach the nominal value for sample sizes of 50 and 100.  Compared to the results presented in Table \ref{water_maze_sim_results}, the type I error rates are consistent across all scenarios and are much closer to the nominal value of 0.05.         

\begin{table}
\caption{Simulated Type I error rates using the LRT comparing two Dirichlet populations for various balanced sample sizes. The $\alpha$ parameters for the Dirichlet distributions are given in the first column. The first row of $\alpha$ parameters are the MLEs from the Maugard study. Error rates were estimated using 10,000 simulations. \label{water_maze_sim_results2}}
\begin{center}
\begin{tabular}{rrrrr}
  \hline
 $\bm{\alpha}$ & $n=7$ & $n=15$ & $n=50$ & $n=100$ \\ 
  \hline
 (9.8, 6.1, 5.4, 5.9) & .0818 & .0627 & .0512 & .0546 \\ 
   (6.8, 6.8, 6.8, 6.8) & .0824 & .0654 & .0492 & .0495 \\ 
   (9, 6, 6, 6) & .0829 & .0647 & .0534 & .0505 \\ 
  (8, 7, 7, 5) & .0809 & .0657 & .0519 & .0508 \\ 
  (12, 5, 5, 5) & .0798 & .0632 & .0591 & .0543 \\ 
   \hline
\end{tabular}
\end{center}
\end{table}

%\subsection{Type II Error for LRT}
To examine the power of the LRT further,  we conducted a simulation study using the  sample sizes of $n=5,7,10,15,20$ for each group and the parameter estimates $\bm{\hat{\pi}_1}$, $\bm{\hat{\pi}_2}$,  $\hat{A}_1$, and $\hat{A}_2$ given previously.  The power estimates are calculated using 10,000 simulated realizations for the observed effect size for the Maugard data. Results show that the sample size used in the Maugard study ($n=7$), the power is a moderate 0.674. Thus, there is only a 67\%  chance to detect a difference if it exists.  For a sample size of 5, the power decreases to 0.547. It would require a sample size of 10 to obtain power above 0.8 (.827), while sample sizes of 15 and 20 result in a statistical power of .845 and .987, respectively. While the small sample size is certainly part of the story, another part might be an incorrect model for the data. We investigate a more flexible model in the next section.

% Don't really need a table to display these data. Prefer to do it inline.
%\begin{table}
%\caption{Simulated statistical power over varying sample sizes $(n)$ using the LRT comparing two Dirichlet populations using the Maugard data as an observed effect size. Power estimates were obtained using 10,000 simulations. \label{water_maze_sim_power}}
%\begin{center}
%\begin{tabular}{rr}
%  \hline
% $n$ & Power  \\ 
%  \hline
%   5 & .5466 \\ 
%   7 & .6741 \\ 
%  10 & .8274 \\ 
%  15 & .9447 \\ 
%  20 & .9870 \\ 
%   \hline
%\end{tabular}
%\end{center}
%\end{table}

\section{Dealing with Positive Correlations among Components}
 There is one important consideration when applying the Dirichlet distribution to a data set: the theoretical correlations between all pairs of components should be negative. Table \ref{sampcorcomb} shows the pairwise correlation structure between components when all 14 observations from the Maugard data set are used for calculation. The correlation between the components AQ1 and OQ, highlighted in  bold face type, is positive. The correlation is small, and it is possible that the true population correlation between the quadrants is negative; however, it is important to have a scenario for comparisons of compositional data structures where the true population correlation is non-negative. For example, the compositional structures for some microbe (or cell) proportions in metagenomic data are likely to be positively correlated with one another \cite{koslovsky2020, tang2017}, as are components of products taken from a forest \cite{hickey2015}, components within a soil sample \cite{coakley}, and many other fields \cite{aitch2005}.

\begin{table}
\caption{The correlation between pairs of components for the probe test in \cite{maugard} for all 14 observations.  Note that the correlation between AQ1 and OQ is positive, indicating that the DD might not be a good fit for these data. \label{sampcorcomb}}
\begin{center}
\begin{tabular}{rrrrr}
  \hline
 & TQ & AQ1 & OQ & AQ2 \\ 
  \hline
TQ & 1.00 & -0.66 & -0.51 & -0.32 \\ 
  AQ1 & -0.66 & 1.00 & \textbf{0.15} & -0.25 \\ 
  OQ & -0.51 & \textbf{0.15} & 1.00 & -0.27 \\ 
  AQ2 & -0.32 & -0.25 & -0.27 & 1.00 \\ 
   \hline
\end{tabular}
\end{center}
\end{table}
%The restriction that pairwise correlations among the components be nonnegative for a Dirichlet distribution is one of the main reasons that Aitchison cautioned against its use for compositional data. Instead, he developed methodology based on the additive log ratio (ALR) and the centered log ratio (CLR)\cite{aitch1982}. 
%The centered log ratio is given by
%\begin{equation}
%    clr(x) := \left(\ln(x_i) - \frac{1}{D}\sum_{j=1}^D\ln( x_j) \right)_i \label{eq:clr}
%\end{equation}

 Certainly, the log--ratio based tests are amenable to positive correlations between components. We performed a CLR analysis of the Maugard data using the {\texttt compositions} R packages \cite{PackageCompositions} and \texttt{DescTools} \cite{DescTools}. We use the CLR because there is no natural reference group for the water maze data, and the CLR does not require this choice, unlike the ALR transformation. After transformation with the CLR, we employed Hotelling's $T^2$ test \cite{anderson} to test the null hypothesis that $H_0: \bm{\lambda_{1i}}=\bm{\lambda_{2j}}$ for all $i$ and $j, i\neq j$, where $\bm{\lambda}$ represents the CLR transformed $\bm{\pi}$ vectors. The subscripts 1 and 2 indicate group membership (WT and AD). For this test, $T^2$ = 1.6385, $df_1$ = 3, $df_2$ = 10, and the p-value is 0.2423. The conclusion from the CLR test indicates no evidence of a difference for any of the quadrants between the WT and AD groups, which corroborates the result from the LRT based on the Dirichlet distribution.

\subsection{A Two-Sample Test based on the Nested Dirichlet Distribution}

A CLR analysis provides a mathematically sound treatment of compositional data and it is simple to use. The transformation moves compositional data from a support between 0 and 1 to an infinite support, and stabilizes the variance, thus allowing for methods based on the Normal distribution, such as Hotelling's $T^2$ test, to be employed validly. In addition, because Hotelling's $T^2$ test is a multivariate test, it takes correlations among the components into account. Furthermore, the correlations between the transformed components can be positive or negative due to the assumption of Multivariate Normality of Hotelling's $T^2$, and they are not restricted by the value of the mean for each group, as is the case with the Dirichlet distribution. In addition to the assumption of multivariate normality, additional assumptions must be met for Hotelling's $T^2$, and other classical analyses, to be employed validly. One of the assumptions of Hotelling's $T^2$ test is that the covariance matrices for the two groups (WT and AD in this case) are the same.  If the covariance matrices are unequal, Hotelling's $T^2$ test might give misleading results \cite{holloway}. Another situation where many classical multivariate tests fail is when the mean vectors are not from the same population. This might happen when measurements are clustered, for example, if mice are tested in batches rather than in a random fashion \cite{anderson}. 

Recall that the sample correlation matrix shown in Table \ref{sampcorcomb} indicates a small positive correlation ($r = .15$) between AQ1 and OQ, implying that a nested Dirichlet distribution is a better model than the Dirichlet distribution for this data set. The tree that describes the nature of the relationship among the components is unknown, and there is no intuitive hierarchy among the four quadrants. Therefore, we must derive the nesting structure using the sample correlation matrix.  

For four variables, there are 20 possible nesting trees. \cite{null2008} notes that if a component variable $x$ is negatively (positively) correlated with a nesting variable $y$, then $x$ will be negatively (positively) correlated with any variable nested under $y$.  We can use this fact to rule out some trees. Once trees that do not follow this rule (so-called ``impossible'' trees) are eliminated from consideration, the MLEs for the remaining trees were obtained and the value of the log-likelihood using Equation \ref{LogLikereg} was calculated.  We chose the tree corresponding to the largest log-likelihood as the nesting tree that best represents the correlation structure for the data. Figure  \ref{best_tree_mle} shows this tree with MLEs for the $\alpha$ parameters shown along each branch.  The variables that are positively correlated with each other, AQ1 and OQ,  are nested under the same node.  The variables that are negatively correlated with OQ are nested under a different node. This tree makes sense with regards to the sample correlation matrix in Table \ref{sampcorcomb}.

MLEs for the parameters of the finalized nested Dirichlet fit are displayed in Figure \ref{best_tree_mle}.  The subtree composition, $\bm{b_{root}}$, with children $N_1$ and $N_2$ is modeled as $\bm{b_{root}}\sim DD(8.1, 11.2)$.  The subtree composition, $\bm{b_{N_1}}$, with children AQ1 and OQ is modeled as $\bm{b_{N_1}} \sim DD(11.6, 10.3)$.  Lastly, the subtree composition, $\bm{b_{N_2}}$, with children AQ2 and TQ is modeled as $\bm{b_{N_2}} \sim DD(5.6, 9.2)$.  
\begin{table}
\centering
\begin{tabular}{rrrrr}
 \hline
& TQ & AQ1 & OQ & AQ2 \\ 
\hline
TQ &  1.00 & -0.55 & -0.53 & -0.31\\ 
 AQ1 & -0.55 & 1.00 & 0.20 & -0.37 \\ 
OQ & -0.33 & 0.20 & 1.00 & -0.33 \\ 
AQ2 & -0.31 & -0.37 & -0.33 & 1.00 \\ 
\hline
\end{tabular}
\caption{The correlation between pairs of variables for the simulated water maze data.  Compare this matrix with that in Table \ref{sampcorcomb}.  Although the values of the correlations are not the same, they are all ``in the ballpark".  Specifically, the signs of the correlations for the simulated data match that of the sample data.}
\label{simcorwm}
\end{table}
To check how well this model matches the sample data, we simulated 1000 draws independently from each of the three subtrees using the function \texttt{rdirichlet} in the package \texttt{gtools} \cite{gtools}.  The simulated data for each subtree were then used to obtain the simulated values of the four component water maze data.  For example, the simulated values for the AQ1 and OQ components would be obtained by first simulating samples from $\bm{b_{root}}$ and $\bm{b_{N_1}}$.  Simulated AQ1 values would be obtained by multiplying the first component of $\bm{b_{root}}$ to the first component of $\bm{b_{N_1}}$ while OQ2 values would be multiplied using the second component of $\bm{b_{N_1}}$. The sample correlation matrix from the simulated data is in Table \ref{simcorwm}.  All of the correlations from the simulated data are near those in Table \ref{sampcorcomb}.  More importantly, the signs of the correlations from the simulated data match the signs of the correlations of the sample data. The nested Dirichlet model does a fine job of capturing the correlation structure.  
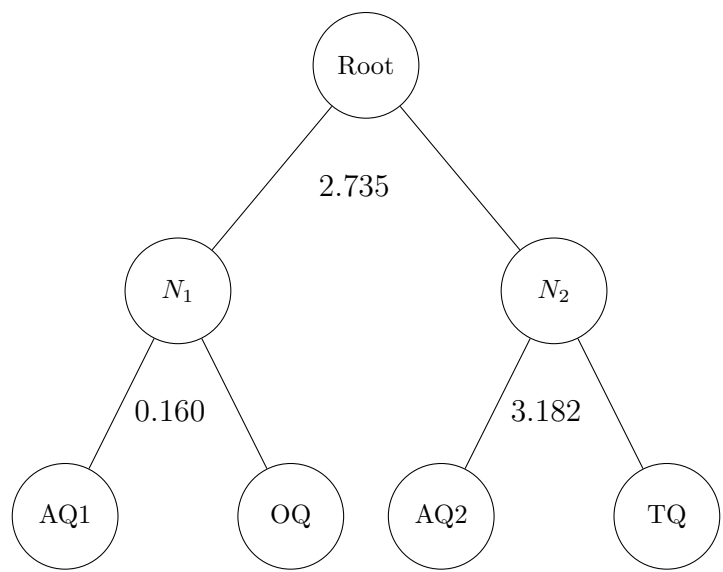
\begin{figure}
\centering
     \begin{tikzpicture}
    \node[circle, draw, minimum size = 4em]{Root}[sibling distance= 5 cm, level distance = 3 cm]
        child {node[circle, draw, minimum size = 4em] {$N_1$} [sibling distance = 3 cm] 
            child{node[circle, draw, minimum size = 4em] {AQ1}
            edge from parent node[right, xshift = 0.05 cm, yshift=-0.1cm] {\large 0.160}}
            child {node[circle, draw, minimum size = 4em] {OQ}}
            edge from parent node[right, xshift = 0.50 cm, yshift=-0.1cm] {\large 2.735}
        }
        child {node[circle, draw, minimum size = 4em] {$N_2$} [sibling distance = 3 cm]
            child {node[circle, draw, minimum size = 4em] {AQ2}
             edge from parent node[right, xshift = 0.05 cm, yshift=-0.1cm] {\large 3.182}}
            child{node[circle, draw, minimum size = 4em] {TQ}}
        };
    \end{tikzpicture}
\caption{The tree structure used for the water maze data with the value of likelihood ratio test statistic for each subtree.}\label{fig:treeLRT}
\end{figure}

%\textcolor{cyan}{Because we have settled on a nested Dirichlet model and found a reasonable structure for the relationships, the two-sample test for the nested Dirichlet model, presented in \cite{turner2013}, termed Layered Dirichlet Modeling (LDM), can be applied. If the means between the two populations are different, then for at least one subtree, the mean of their compositions will be different as well.  A global likelihood ratio test can be constructed with this hypothesis in mind. }
%

Now that we have settled on a nested Dirichlet model and found a reasonable nesting structure for the relationships among components, we introduce a two-sample test to determine if changes in the composition exist. Let the mean vectors for the respective nested Dirichlet populations be denoted as $\bm{\pi_1}$ and $\bm{\pi_2}$.  To test the hypotheses $\text{H}_0: \bm{\pi_1}=\bm{\pi_2}$ we can consider an equivalent set of hypotheses based on the fact that, if the means between the two populations are different, then for at least one subtree, the mean of their compositions will be different as well.  To formally write the hypotheses we must first consider the following notation.

Let $N_0$ be defined as the root node, and let $N_i$ for $i=1,2,...,l$ correspond to the remaining internal nodes within the tree.
The mean vectors for each of the $i=0,1,...,l$ subtrees by treatment group are denoted as $\bm{\pi_{1,N_i}}$ and $\bm{\pi_{2,N_i}}$ respectively.  The equivalent test for the difference in mean compositions between two groups of the nested Dirichlet can be expressed as
\begin{equation}
\label{eqn:nddtest}
\begin{split}
\text{H}_0: & \hspace{0.1in} \bm{\pi_{1,N_i}} = \bm{\pi_{2,N_i}} \hspace{0.1in} \text{For all $i$}\\
\text{H}_1: & \hspace{0.1in} \bm{\pi_{1,N_i}}\neq\bm{\pi_{2,N_i}} \hspace{0.1in} \text{For at least one $i$}
\end{split}  
\end{equation}
A likelihood ratio test can easily be constructed since the subtree compositions are independent and distributed Dirichlet. Let $\Lambda_i$ be the likelihood ratio test statistic for the $i^{\text{th}}$ subtree  $ i =0,1,...,l$.  The test statistic to determine if there are differences between the mean vectors for the two groups of mice is $\Lambda_{\text{overall}}=\sum^{l}_{i=0} \Lambda_i$, which follows a $\chi^2$ distribution with degrees of freedom $\sum_{i=0}^{l} (k_i-1)$, where $k_i$ is the number of variables in the $i^{\text{th}}$ subtree. Note that the test statistic is just the sum of the individual likelihood ratio test statistics for each of the independent Dirichlet distributions. This feature is unique to the nested Dirichlet test, and it allows the analyst to examine the individual LRT statistics to see which subtree contributed most to the overall test statistic and gain insight to which components within the composition show the most change. Also note that the LRT for the Dirichlet setting is a special case of the LRT under the nested Dirichlet model. This is true because $\Lambda_{overall}$ reduces to $\Lambda_0$, which corresponds to all components of the composition being nested under the root node with no additional nesting structure. 

 In our specific case of Nested Dirichlet for the Maugard data, we have 3 nodes $N_0,N_1,N_2$ with $k_i = 2$ for all $i$. Figure \ref{fig:treeLRT} gives the likelihood ratio test statistic for each subtree in the example.  Summing the likelihood ratios for the individual Dirichlet distributions, we obtain $\Lambda_{\text{overall}} = 2.735 + 0.16 + 3.18 = 6.08$.  The two larger LRT statistics are for the root node, $N_0$, and $N_2$ which both contain compositional information involving the TQ component, which was observed to have the biggest shift in mean via summary statistics and ternary plots. The overall test statistic follows a $\chi^2$ distribution with 3 degrees of freedom.  The p-value associated with this test statistic is 0.108.  Compare this p-value to 0.067 and 0.2423, which are the p-values using a non-nested Dirichlet distribution and the CLR analysis, respectively.  In any case, the conclusion remains the same for $\alpha=0.05$; there is not enough evidence to conclude a difference between the two groups of mice. The result of  hypothesis tests appropriate for composition data, whether the CLR or our statistical tests for the Dirichlet and Nested Dirichlet models, disagree with the initial analysis conclusions of \cite{maugard}.
 %\textcolor{cyan}{This result is similar to the CLR analysis and the previous analysis we completed with these data. These results do not support evidence for a difference in time spent within the various parts of the water maze.}

\subsection{Confidence Intervals for Pairwise Comparisons}
 After rejecting the null hypothesis in either \eqref{mice_hypo} or \eqref{eqn:nddtest} , it will be of interest to determine what pairs of components contribute to the overall rejection as well as estimate the magnitude of the difference in the mean vectors to gain further practical insights.  Confidence intervals for the difference in means for each of the variables within a Dirichlet or nested Dirichlet composition are derived using the large sample properties of maximum likelihood estimators.  While the estimates for the mean difference are easy to compute, the standard errors require some additional care and are increasingly cumbersome when moving from the Dirichlet to the nested Dirichlet case.  The derivation for these intervals for both the Dirichlet and nested Dirichlet cases are provided in the appendix. We strongly emphasize, as is done with traditional multivariate analysis and univariate one-way ANOVA analysis, that posthoc exploration of individual comparisons should only be explored after the null hypothesis for the overall LRT is rejected. Our presentation of the intervals for the Maugard water maze data is done to provide example computations of our derived interval formulas rather than to draw conclusions.  

As a numeric example, Table \ref{DirCI_WaterMaze} provides estimates, standard errors, and confidence intervals for the difference in means comparing WT and AD for the four components in the composition. Since there are four intervals being produced, the confidence levels were adjusted for multiple comparisons using a Bonferroni correction. 
\begin{table}
\caption{Parameter estimates and confidence intervals for comparing AD vs WT mean by each variable in the composition using the Dirichlet model.  Mean and difference estimates are provided with their corresponding standard errors in (). Confidence intervals were adjusted for multiple comparisons using the classic Bonferroni procedure. \label{DirCI_WaterMaze}}
\begin{center}
\begin{tabular}{rrrrrr}
  \hline
  Component & $\hat{\bm{\pi}}_{AD}$ & $\hat{\bm{\pi}}_{WT}$  & $\hat{\bm{\pi}}_{AD}-\hat{\bm{\pi}}_{WT}$  & $(1-.05/4)100\%$ CI \\ 
  \hline
   TQ  & .301 (.026) & .423 (.035) & -.121 (.044) & (-.231,-.012) \\ 
   AQ1 & .255 (.025) & .194 (.028) & .060 (.037)  & (-.033,.153)  \\ 
   OQ  & .216 (.023) & .181 (.027) & .035 (.036)  & (-.054,.124)  \\ 
   AQ2 & .228 (.024) & .202 (.028) & .026 (.037)  & (-.066, .118) \\ 
   \hline
\end{tabular}
\end{center}
\end{table}
The confidence intervals for AQ1, OQ, and AQ2 all contain 0 while the interval for TQ does not, indicating that a change in mean is probably between WT and AD groups for the target quadrant. This is consistent with our observations throughout our exploratory analysis that a shift in the means for TQ is plausible. However, we emphasize here again the importance of exploring multiple interval comparisons only after a rejected overall LRT in practice.  

Table \ref{NDirCI_WaterMaze} provides estimates, standard errors, and confidence intervals for the difference in means comparing WT and AD for the four components in the composition using the proposed NDD model. As with the Dirichlet setting, the confidence levels were adjusted for multiple comparisons using a Bonferroni correction. 

\begin{table}
\caption{Parameter estimates and confidence intervals for comparing AD vs WT mean by each variable in the composition using the Nested Dirichlet model.  Mean and difference estimates are provided with their corresponding standard errors in (). Confidence intervals were adjusted for multiple comparisons using the classic Bonferroni procedure. \label{NDirCI_WaterMaze}}
\begin{center}
\begin{tabular}{rrrrrr}
  \hline
  Component & $\hat{\bm{p}}_{AD}$ & $\hat{\bm{p}}_{WT}$  & $\hat{\bm{p}}_{AD}-\hat{\bm{p}}_{WT}$  & $(1-.05/4)100\%$ CI \\ 
  \hline
   TQ  & .303 (.028) & .425 (.038) & -.122 (.048) & (-.241,-.004) \\ 
   AQ1 & .253 (.023) & .194 (.028) & .059 (.036)  & (-.030,.148)  \\ 
   OQ  & .215 (.021) & .180 (.026) & .035 (.034)  & (-.050,.119)  \\ 
   AQ2 & .230 (.026) & .201 (.028) & .029 (.038)  & (-.067, .124) \\ 
   \hline
\end{tabular}
\end{center}
\end{table}

The confidence intervals for the Nested Dirichlet model are quite similar to those computed using the Dirichlet model.  While the interval for TQ still doesn't contain 0, the interval is wider and closer to including $0$ than the Dirichlet model's interval estimates and is consistent with the more conservative LRT for the nested Dirichlet model.  The close similarity between the two models also suggests that the added flexibility of the Nested Dirichlet was not really utilized. This is  evident from a very small positive correlation in just one of the pairs.  Had the positive correlation been stronger and the correlations less consistent with a standard Dirichlet, it is likely that a greater discrepancy in the SE and the interval widths would have been observed. 

\section{Transparancy and Openness}

In the interest of computational reproducibility, we have also provided a GitHub page containing R scripts that produce all results, tables, and figures in the main document and in supplementary materials (https://github.com/Bianca-Luedeker/WaterMaze).  The code provides two key benefits. The first is a demonstration using a real data analysis that offers clear examples of how to use the methodology described.  Secondly, the simulation scripts allow for researchers to effectively design future water maze experiments by considering how effect size and sample size impact the power of the two-sample Dirichlet and nested Dirichlet hypothesis tests. 

\section{Summary and Conclusion}
While data made up of components where the proportions of responses within all components must sum to 1 are plentiful, the correct analyses of these data sets are not. In particular, previous data analyses of data from the Morris water maze experiment have focused on the comparison of time spent in the target quadrant as a univariate comparison between groups. In fact, of the 25 articles published using Morris Water Maze experiments from 2018 and 2019, 24 of them have used classical tests such as t-tests,  ANOVA, and ANCOVA \cite{maugard}. Such an analysis fails to recognize the inherent correlation between times spent in each quadrant. Furthermore, previous methods do not recognize that the proportion of time spent in each quadrant should sum to 1.  In the specific case of two sample comparisons, the authors of \cite{maugard} take a step in the right direction by analyzing a set of Morris Water Maze data using a Dirichlet distribution rather than employing a component-wise analysis based on a t-test or a nonparametric equivalent.  The analysis needs to be taken two more steps forward. First, careful consideration when aligning the research question of interest to a specific null and alternative hypothesis in terms of Dirichlet parameters is critical as illustrated by our simulation studies of their procedure. Secondly, while the Dirichlet distribution is a model appropriate for compositions, its distributional properties are quite rigid both in terms of its mean-variance relationship and correlation structure.  Applying two sample comparison methodology using compositional models that are more flexible in fitting characteristics of real data is something to consider.

% These results were further extended to Dirichlet and nested Dirichlet distributions when there are three or more populations being compared \cite{luedeker}. 
To address these issues, we have presented statistically sound methodology to compare mean vectors of two populations when compositional data is assumed to follow Dirichlet and nested Dirichlet distributions.  These advancements offer additional tools to analyze compositional data such as Morris water maze experiments and complements previous compositional approaches based on transformations using log ratios.  While the Dirichlet has been criticized for being too restrictive, conducting statistical inference for the nested Dirichlet distribution provides additional flexibility to model more real world compositions.  The simplicity of the logratio method is its main advantage; however, confidence intervals for the difference in means for each component are obtained on the transformed scale.  We are not aware of any method that allows for these intervals to be back transformed onto the original scale.  If the Dirichlet or nested Dirichlet model is a reasonable assumption, our methodology provides confidence intervals for the difference in means for each component directly; thus leading to easily interpretable confidence intervals on the same scale as the original data. 

The correlation structure of the data analyzed by \cite{maugard} indicated that a nested Dirichlet distribution is a more appropriate model than a Dirichlet distribution. The correlations between proportions of time spent in all four quadrants from the Maugard water maze data are both positive and negative, while correlations among components in a Dirichlet distribution are constrained to be negative. However, the positive correlations found in Table  \ref{sampcorcomb} were small (only $0.15$), which might be small enough to be negligible with regard to the fit of the Dirichlet distribution to the data. Our definition of ``small'' is arbitrary. The effect of ignoring small positive correlations on the performance of the likelihood ratio test or on the fit of the Dirichlet distribution were partially addressed in \cite{luedeker} and are a subject for continuing research.  Both of our proposed LRT approaches agreed with the CLR transformation analysis. All three analyses (using CLR, the Dirichlet distribution and the nested Dirichlet distribution) contradicted the conclusion of the original analysis method conducted by \cite{maugard}. 

Conducting statistical inference using the nested Dirichlet distribution creates additional considerations.  The NDD relies on a fixed tree structure that must be determined in advance, which might seem like an unnecessary added step. However, for complex correlation structures, such a step is critical as the structure provides added flexibility to model the characteristics of a potentially complex compositional data set. The Maugard data has no natural tree structure; therefore, we generated a structure from the sample correlation matrix and chose a tree based on a theoretical result of the NDD that correlation between components that are nested under the same internal node must have the same sign \cite{null2009}. The tree structure in Figure \ref{fig:treeLRT} was one of several trees that are a ``reasonable fit''. The criteria used to choose this tree was maximum log-likelihood, which is possible to apply in a brute force way because of the small number of possible trees for these data. In the case where there are many possible trees, it is vital to have a data driven algorithm to find such a structure rather than doing calculations ``by hand". The details of this algorithm and its extension have been previously discussed, and are beyond the scope of this paper \cite{turner2013, luedeker}.
%The nesting structure in this case is based entirely on the correlation structure among the variables, not any physical real life relationship among the variables. It would be a good idea to have a data-driven algorithm to find such a structure rather than choosing an appropriate tree out of a range of feasible candidates. That way the tree would not be determined by human judgment.  % 

It is also of interest to examine the robustness of the LRT procedures when the tree is incorrectly specified to varying degrees.  This is particularly important when the nesting structure is not apparent or natural and must be derived from the correlation structure among the components.  Choosing a tree based on a fit criterion such as the log-likelihood has a strong mathematical base, while creating an arbitrary tree would be more problematic.  The chosen tree should match the correlation structure in the data \cite{null2009}. An investigation  of the effect of tree choice on test outcome could lead to a simulation--based test for the differences in components and provide additional rules and guidelines for particular scenarios.

The methods based on the DD and NDD have a strong theoretical basis. The NDD, in particular, allows extensions for complex correlation structures among the components, flexible variance values independent of the mean, and presence of positive correlations among components.  Additionally, the LRT tests provided have a practical interpretation of the relationships among components within a compositional data set. It is easy to determine which component contributes the most to the magnitude of the overall test statistic based on the magnitude of the likelihood ratios for each level of the tree.  Together with methods based on log ratios, they provide valid methods for examining differences among the mean component vectors for two populations. 

\bibliographystyle{plain}
\bibliography{thesis_bibV1}

\vfill

\newpage

\section{Supplementary Material: Additional Theoretical Details}
Here we give the mathematical details of the derivation of post-hoc confidence intervals for both the DD and NDD cases. The first derivation provides standard error estimates for the mean of a standard Dirichlet distribution for a single sample.  From this result, a confidence interval for the difference in means is easily obtained.  The second derivation utilizes the delta method to obtain the standard errors for the mean of a nested Dirichlet distribution for the one sample case. An extension of the result for the difference in two means is easily derived from the one sample result.  

\subsection{Confidence Intervals for Difference in Means (Dirichlet Case)}

Under certain regularity conditions that the Dirichlet distribution satisfies, standard errors for maximum likelihood estimators can be obtained for large samples by deriving Fisher's information.  In general, let $L(\bm{\theta|data)}$ be the log-likelihood function for a sample of independent observations where $\bm{\theta}$ is a vector of parameters.  Fishers information, $I(\bm{\theta})$, is the negative expected value of the matrix of partial derivatives of  $L(\bm{\theta}|data)$.  The $(i,j)^{th}$ element of Fisher information is formally defined as
\begin{equation}
    I(\bm{\theta})_{ij}=-E \left( \frac{\partial^2 L(\bm{\theta}|data)}{\partial \theta_i \partial \theta_j} \right)
\end{equation}

In the case of the Dirichlet distribution with $k$ components for a single population, the parameter vector $\bm{\theta}=(\bm{\pi},A)$ and is defined element-wise as $(\pi_1,\pi_2,...,\pi_{k-1},A)$. It should be noted that $\pi_k$ is not included since the mean vector $\bm{\pi}$ is constrained to sum to one. The final mean value, $\pi_k$, serves as the constrained parameter written completely with respect to the previous terms, $1-\Sigma_{j=1}^{k-1} \pi_j$. Let $L(\pi,A|data)$ be log-likelihood for a single sample of $n$ from a Dirichlet distribution defined in Equation \ref{eq:dd}.  

We will denote Fisher's information as the $k\times k$ matrix $I(\bm{\pi},A)$.  The first $k-1$ rows and columns correspond to the partial derivatives of $L(\bm{\pi},A|\text{data})$ with respect to the first $k-1$ elements in $\bm{\pi}$. The remaining off-diagonal entries correspond to the partial derivatives with respect to each of the first $k-1$ elements in $\bm{\pi}$ and $A$.  The final $(k,k)$ entry corresponds to differentiating with respect to $A$ twice. The first partial derivatives are, respectively
\begin{equation}
    \begin{matrix*}[l]
    \frac{\partial L(\bm{\pi},A|\text{data})}{\partial \pi_j} =-nA\Psi(A\pi_j)+nA\Psi(A\pi_k)+ nA\mean{\text{log}x_j}-nA\mean{\text{log}x_k}  \\
    \frac{\partial L(\bm{\pi},A|\text{data})}{\partial A} = n\Psi(A)-n\Sigma_{j=1}^k \pi_j\Psi(A\pi_j)+n\Sigma_{j=1}^k \pi_j\mean{\text{log}x_j}.
    \end{matrix*}
\end{equation}
where $\Psi$ is the digamma function defined as $\frac{\partial}{\partial x} log \Gamma(x)$.

The first $k-1$ diagonal entries of $I(\bm{\theta},A)$ are
\begin{equation}
\begin{split}
    I(\bm{\pi},A)_{jj} &= -E \left( \frac{\partial^2 L(\bm{\pi},A|\text{data})}{\partial \pi_j \partial \pi_j} \right) \\ &= -E(-nA^2 \Psi'(A\pi_j)-nA^2\Psi'(A\pi_k))\\
            &= nA^2 \Psi'(A\pi_j)+nA^2\Psi'(A\pi_k).
\end{split}
\end{equation}
for $j=1,...,k$.

Similarly, the last diagonal entry is
\begin{equation}
\begin{split}
    I(\bm{\pi},A)_{kk} &= -E \left( \frac{\partial^2 L(\bm{\pi},A|\text{data})}{\partial A \partial A} \right) \\ &= -E(n\Psi'(A)-n\Sigma_{j=1}^k\pi_j^2\Psi'(A\pi_j)  )
  \\ &= n\Sigma_{j=1}^k\pi_j^2\Psi'(A\pi_j)-n\Psi'(A).
\end{split}
\end{equation}

The off diagonals for the first $k-1$ rows and columns of $I(\bm{\pi},A)$ are
\begin{equation}
\begin{split}
    I(\bm{\pi},A)_{j,l} &= -E \left( \frac{\partial^2 L(\bm{\pi},A|\text{data})}{\partial \pi_j \partial \pi_l} \right) \\ &= -E(-nA^2\Psi'(A\pi_k))\\
            &= nA^2\Psi'(A\pi_k).
\end{split}
\end{equation}
for $j \neq l$, $j=1,...,k-1$, and $l=1,...,k-1$.

The previous derivations required calculating the expected value of a constant because the terms involving the sample data, $x_{ij}$, vanish upon differentiation. The remaining off diagonals along the last row and column are derived similarly, but the expectation involves functions of $x_{ij}$; therefore, additional care is needed.   Taking the expected value of partial derivatives between $\pi_j$ and $A$, we have
\begin{equation}
\label{PartialApi}
\begin{split}
 I(\bm{\pi},A)_{j,l} &=
-E \left( \frac{\partial^2 L(\bm{\pi},A|\text{data})}{\partial \pi_j \ \partial A} \right) \\ &=
-E(-nA\pi_j\Psi'(A\pi_j)+nA\pi_k\Psi'(A\pi_k)-n\Psi(A\pi_j)+n\Psi(A\pi_k)+n\mean{\text{log}x_j}-n\mean{\text{log}x_k}  ) \\
&= nA\pi_j\Psi'(A\pi_j)+nA\pi_k\Psi'(A\pi_k)+n\Psi(A\pi_j)-n\Psi(A\pi_k)-nE(\mean{\text{log}x_j})+nE(\mean{\text{log}x_k}).
\end{split}
\end{equation}

Due to properties of exponential families, we know that $E(log(x_{ij}))=\Psi(A\pi_j)-\Psi(A)$ \cite{ng2011}. Upon substitution of the expected values in Equation \ref{PartialApi}, the last four terms cancel, leaving the final result
\begin{equation}I(\bm{\pi},A)_{j,l}=nA\pi_j\Psi'(A\pi_j)+nA\pi_k\Psi'(A\pi_k).
\end{equation}
for $j=1$ while $l=1,...,k-1$ and $l=1$ while $j=1,...,k-1$.

Let $\hat{\bm{\pi}}$ and $\hat{A}$ denote the maximum likelihood estimators for  $L(\bm{\pi},A|data)$ and consider the random vector $(\hat{\pi}_1,\hat{\pi}_2,...,\hat{\pi}_{k-1},\hat{A})^T$. For large sample sizes, the sampling distribution of the random vector is multivariate normal with mean vector $(\pi_{1},\pi_{2},...,\pi_{k-1},A)^T$ and variance-covariance matrix 
 \begin{equation}
 \Sigma=I(\pi,A)^{-1}.
 \end{equation}
The standard errors for the mean estimates correspond to the square root of the first $k-1$ diagonals of $\Sigma$.  For a given data set, we can estimate the standard errors by substituting $\hat{\pi}$ and $\hat{A}$ into $I(\pi,A)^{-1}$. The variance of the final mean estimate $\hat{\pi}_k$ can be computed from the others by noting

\begin{equation}
\begin{split}
    Var(\hat{\pi}_k) &= Var(1-\Sigma_{j=1}^{k-1} \hat{\pi}_j) \\
                     &= Var(\Sigma_{j=1}^{k-1} \hat{\pi}_j) \\
                     &= \Sigma_{j=1}^{k-1} \Sigma_{l=1}^{k-1} I(\bm{\pi},A)_{jl}^{-1}
\end{split}    
\end{equation}

%   \begin{equation}
%\left(\begin{array}{c}
%\hat{\pi}_{1}\\
%\hat{\pi}_{2}\\
%\vdots \\
%\hat{\pi}_{k-1}\\
%\hat{A}
%\end{array}\right)\sim MVN\left(\left(\begin{array}{c}
%\pi_{1}\\
%\pi_{2}\\
%\vdots\\
%\pi_{k-1}\\
%A
%\end{array}\right),I(\pi,A)^{-1}\right).\label{eq:MLE_MVN}
%\end{equation}
 
To obtain a confidence interval for the mean difference for a given component, let $\hat{\bm{\pi}}_1$, $\hat{\bm{\pi}}_2$, $\hat{A}_1$, and $\hat{A}_2$ be the respective MLEs for the mean and precision parameters for treatment 1 and 2 obtained from maximizing the log-likelihood in Equation \ref{LogLikereg}.  Assuming independent samples, a $(1-\alpha)100\%$ confidence interval for the difference in means for the $j^{th}$ variable in the composition, $\pi_{1j}-\pi_{2j}$, is 
\begin{equation}
    \hat{\pi}_{1j}-\hat{\pi}_{2j}\pm z_{\frac{\alpha}{2}}\sqrt{Var(\hat{\pi}_{1j})+Var(\hat{\pi}_{2j})}
\end{equation}
where $Var(\hat{\pi}_{1j})$ and $Var(\hat{\pi}_{2j})$ are obtained from $I(\hat{\pi}_1,\hat{A}_1)^{-1}$ and $I(\hat{\pi}_2,\hat{A}_2)^{-1}$ respectively.

\subsection{Confidence Intervals for Difference in Means (nested Dirichlet Case)}
Under the nested Dirichlet distribution, parameter estimation is conducted by estimating the mean and precision using a standard Dirichlet distribution for each of the designated subtrees defined by the internal nodes.  Let $N_0$ be defined as the root node, and let $N_i$ for $i=1,2,...,l$ correspond to the remaining internal nodes within the tree. Denote the parameter estimates for each subtree as $\hat{\bm{\pi}}_{N_i}$ and $\hat{A}_{N_i}$ for $i=0,1,...,l$.  For any given component $X_j$ within the nested Dirichlet, the maximum likelihood estimate for its mean, $\hat{p}_j$, is the product of the elements within each $\hat{\bm{\pi}}_{N_i}$ that correspond to the path of the tree from the root down to terminal node denoted as $X_j$. For the example in Figure \ref{fig:toy_tree}, $\hat{p}_2$ is the product of the first entry of $\hat{\bm{\pi}}_{N_0}$ and the second entry of $\hat{\bm{\pi}}_{N_1}$

In general, the mean estimate for an individual component within the nested Dirichlet is always the product of mean estimates from the subtrees following the root node down to its corresponding terminal node.  To simplify mathematical notation, let $\hat{\bm{\pi}}(X_j)$ be the vector of subtree mean estimates required in the product to obtain the estimate for $X_j$ and let $\bm{\pi}(X_j)$ denote the true parameter values for which $\hat{\bm{\pi}}(X_j)$ estimates.  Due to the independence of the subtrees and the fact that at most one estimate is used from each subtree, large sample theory dictates that the random vector $\hat{\bm{\pi}}(X_j)$ follows a multivariate normal distribution with mean vector $\bm{\pi}(X_j)$ and covariance matrix denoted $\Sigma(X_j)$ which is diagonal with elements defined by the corresponding variance estimates for each element of $\hat{\bm{\pi}}(X_j)$ obtained by their corresponding diagonal entry within $I({\bm{\pi}}_{N_i},{A}_{N_i})^{-1}$.  

For any vector $\bm{y}$ of length $k$, define the function $h$ to be the product of all entries in $\bm{y}$, $h(\bm{y})=\prod_{i=1}^k y_i$. The corresponding MLE for the mean of $X_j$ of the nested Dirichlet can be written as $\hat{p}_j=h(\hat{\bm{\pi}}(X_j))$. Based on the multivariate delta method, $\hat{p}_j$ follows a normal distribution with mean $p_j=h(\bm{\pi}(X_j))$ and variance 
\begin{equation}
\label{delta}
    Var(\hat{p}_j)=\nabla h(\bm{\pi}(X_j))^{T} \Sigma(X_j) \nabla h(\bm{\pi}(X_j))
\end{equation}
where $\nabla h(\bm{\pi}(X_j))$ is the gradient vector.

To obtain a confidence interval for the mean difference for a given component of the nested Dirichlet, let $\hat{\bm{p}}_1$, $\hat{\bm{p}}_2$ be the vector of mean estimates for treatment group 1 and 2 obtained via computing $\hat{p}_j=h(\hat{\bm{\pi}}(X_j))$ for  $j=1,...,k$.  Assuming independent samples, a $(1-\alpha)100\%$ confidence interval for the difference in means for the $j^{th}$ variable in the composition, $p_{1j}-p_{2j}$, is 
\begin{equation}
    \hat{p}_{1j}-\hat{p}_{2j}\pm z_{\frac{\alpha}{2}}\sqrt{Var(\hat{p}_{1j})+Var(\hat{p}_{2j})}
\end{equation}
where $Var(\hat{p}_{1j})$ and $Var(\hat{p}_{2j})$ are obtained from Equation \eqref{delta} replacing the parameters with their corresponding maximum likelihood estimates.

%\bigskip
%\begin{center}
%{\large\bf SUPPLEMENTARY MATERIAL}
%\end{center}

%\begin{description}

%\item[Title:] Brief description. (file type)

%\item[R-package for  MYNEW routine:] R-package (name of package) containing code to perform the diagnostic methods described in the article. The package also contains all datasets used as examples in the article. (GNU zipped tar file)

%\item[HIV data set:] Data set used in the illustration of MYNEW method in Section~ 3.2. (.txt file)

%\end{description}

\end{document}